\documentclass[runningheads, english, a4paper]{llncs}
\usepackage[utf8]{inputenc}
\usepackage{babel}
\usepackage{graphicx}
\usepackage{amsmath, amsfonts}
\usepackage{enumerate}
\usepackage{bm}
\usepackage{rotating}
\usepackage{tikz}
\usepackage{soul}
\usepackage{multirow}
\usepackage{tabularx}
\usetikzlibrary{arrows,fit,backgrounds,decorations.pathmorphing,automata}
\usepackage{cleveref}
\crefname{section}{Sect.}{Sects.}\Crefname{section}{Section}{Sections}
\crefname{figure}{Fig.}{Figs.}\Crefname{figure}{Figure}{Figures}
\crefname{table}{Tab.}{Tabs.}\Crefname{table}{Table}{Tables}
\usepackage{enumitem}

\newcommand{\psosp}{PSOPP} %PSO$_{\text{MSPP}

\newcommand{\neighborhood}[1]{\mathcal{N}_{#1}}

\newcommand{\minPartitionSize}{s_{\mathit{min}}}
\newcommand{\maxPartitionSize}{s_{\mathit{max}}}
\newcommand{\minNbPartitions}{n_{\mathit{min}}}
\newcommand{\maxNbPartitions}{n_{\mathit{max}}}

\newcommand{\partitioningA}{\mathcal{P}}

\newcommand{\resultingPartitioning}[1]{{#1}^\prime}

\newcommand{\partitionA}{K}
\newcommand{\partitionB}{L}
\newcommand{\partitionC}{M}
\newcommand{\partitionD}{N}

\newcommand{\fitnessSym}{f}
\newcommand{\fitness}[1]{\fitnessSym(#1)}

\newcommand{\aParticle}{\Pi_i}
\newcommand{\bParticle}{\Pi_j}

\newcommand{\best}{\mathcal{B}}
\newcommand{\globalBest}{\best}
\newcommand{\bestParticle}[1]{\best_{#1}}
\newcommand{\bestNeighborhood}[1]{\best_{\neighborhood{#1}}}

\newcommand{\framework}{IsoTeSO}

\allowdisplaybreaks
\useshorthands{"}
\defineshorthand{"=}{-\allowhyphens}

\title{An Approach for Isolated Testing of Self-Organization Algorithms}
\titlerunning{An Approach for Isolated Testing of SO Algorithms}
\author{Benedikt Eberhardinger \and Gerrit Anders \and Hella Seebach  \and Florian Siefert \and Alexander Knapp \and Wolfgang Reif}
\authorrunning{B.~Eberhardinger et al.}
\institute{Institute for Software \& Systems Engineering, University of Augsburg, Germany\\
\email{$\{$eberhardinger$,\;$anders$,\;$seebach$,\;$siefert$,\;$knapp$,\;$reif$\}$@isse.de}}

\begin{document}

\maketitle

\begin{abstract}
%The main advantages of self-organizing, adaptive systems (SOAS) are their flexibility and robustness in ever-changing environments. 
%However, their characteristics make it hard to test them adequately.
We provide a systematic approach for testing self"=organization (SO) algorithms.
The main challenges for such a testing domain are the strongly ramified state space, the possible error masking, the interleaving of mechanisms, and the oracle problem resulting from the main characteristics of SO algorithms: their inherent non-deterministic behavior on the one hand, and their dynamic environment on the other.
A key to success for our SO algorithm testing framework is automation, since it is rarely possible to cope with the ramified state space manually.
The test automation is based on a model-based testing approach where probabilistic environment profiles are used to derive test cases that are performed and evaluated on isolated SO algorithms.
Besides isolation, we are able to achieve representative test results with respect to a specific application. 
For illustration purposes, we apply the concepts of our framework to partitioning-based SO algorithms and provide an evaluation in the context of an existing smart-grid application.

\keywords{Self"=organizing Systems, Adaptive Systems, Self"=organization Algorithms, Software Engineering, Quality Assurance, Software Test}
\end{abstract}

%\comment{we test the controller part of the CEI}
%\comment{Clarify: Why do we generate multiple test sequences per system configuration?}

\section{Introduction}\label{sec:intro}
%Self-organization and adaptation enable systems to fulfill their goals in an ever-changing environment by reorganizing themselves during run-time.
%However, this leads to challenges in engineering those systems. 
%It is especially hard to gain assurances for those systems, but even more necessary to take appropriate measures.
The established quality assurance measure of testing aims at systematically covering possible paths through the system with the goal of finding failures.
In this paper we provide a systematic approach for testing self-organization (SO) algorithms. 
This contribution is an elaboration and extension of the first vision of a framework for testing SO algorithms shown in~\cite{eberhardinger2015framework} and is integrated into our overall research road map aiming at testing self-organizing, adaptive systems (SOAS); the vision of this overall approach is outlined in~\cite{Eberhardinger2014TeSOS}.

%Our contribution---which is an elaboration and extensive extension of the first vision of a framework for testing SO algorithms shown in~\cite{eberhardinger2015framework}---is to provide a systematic approach for testing self-organization algorithms; the resulting framework is embedded in our overall approach for testing self-organizing, adaptive systems, the vision of this overall approach is outlined in~\cite{Eberhardinger2014TeSOS}.
%Yet, testing self-organizing, adaptive systems remains a open challenge since their system property make it rarely possible to cover all paths and to decide whether or not a failure occurs.
%This is due to the huge state space to be covered and the fact that error masking is very likely to occur, as we will discuss in more detail later on.
%The latter one addresses the so called oracle problem which encompass the problem of deciding about correct and incorrect states. 
The properties of the SO algorithms like inherent non-deterministic behavior, an ever-changing environment, a high number of interacting components, and interleaving operations make it hard to achieve a systematic testing approach for SO algorithms.
A key to success is automation, since manually it is rarely possible to cope with the high number of demanded test cases (that are necessary as a result of the huge state space).
%However, to be able to automate the process of testing SO algorithms it is necessary to select the test cases to be generated appropriately since not every test case could be executed.
%Furthermore, test cases need to be executed in a controlled environment since faulty behavior of one SO algorithm might be compensated by another mechanisms of the system what makes is rarely possible to observe a failure. 
%Observing failures is the last step of the desired procedure which leads to the well know oracle problem~\cite{Binder1999}, stating that is hard to automatically decide weather a situation is faulty or not. 
However, the automation of testing SO algorithms is faced by the complexity of the system class requiring techniques to cope with the following key challenges: 

\begin{description}[font=\rmfamily\itshape\bfseries,
                    leftmargin=0pt,
                    itemsep=4pt,
                    labelsep=1em,
                    style=standard]
	\item[C-ErrorMask] SO algorithms---and in general SOAS---are designed to be robust and flexible under ever-changing environmental conditions. As a side-effect of these properties an SO algorithm itself, other SO algorithms, or adaptation mechanisms might cover the tracks of possible failures that should be revealed during testing the SO algorithms. Thus, faulty behavior of one SO algorithm could be compensated by another mechanism masking the failure. 
	
	For instance, assume an erroneous SO algorithm that returns wrong or inappropriate system structures as a result to the controlled components. This fault then could be masked by an adaption mechanism of the components that compensates the wrong system structure by a high and costly amount of adaptation. The SO algorithm would encompass a fault, but no failure is visible at first glance since the adaptation mechanism masks it by keeping the system alive. 
	
	\item[C-Isolate]
	SO algorithms are based on interaction with the system's components. In the majority of cases, several different SO algorithms are incorporated; their overlap of interaction with the components leads to so-called interleaved feedback loops. These are challenging in testing, because it is hard to get dedicated results for a single SO algorithm. To address this challenge, there is a need for isolated testing of single SO algorithms. 
	
	As an example of this challenge, assume an SO algorithm that forms organizational structures, e.g., by partitioning the system's components.
	A further algorithm, however, performs its calculations for parametrization of the components based on this structure, e.g., for forming an evenly distributed output for each partition.
	These two algorithms are interleaved and a dedicated test result for the algorithm performing its calculation on the structure is only possible if they are isolated, since the results depend on the results of the first and vice versa.
	However, this isolation is hard to achieve due to the high dependencies between the two algorithms. 
	
	\item[C-Oracle]
	The oracle problem is a well-known challenge for all testing endeavors~\cite{Binder1999,hierons:tse:2011}. 
	However, the properties of SO algorithms increase this problem: For classical testing the conditions of execution for the system under test (SuT) as well as the concrete requirements are known. Let us call these facts the ``known-knowns''.\footnote{The classification of known-knowns, known-unknowns, and unknown-unknowns is borrowed from United States Secretary of Defense Donald Rumsfeld's response given to a question at a U.S. Department of Defense news briefing on February 12, 2002.}
	For SO algorithms under test (SOuT) we know that there are unknown conditions of execution where we can hardly decide a priori, i.e., at design-time, whether a state is correct or not; we call these conditions the ``known-unknowns''.
	Moreover, for the SOuT there might even be situations we are not aware of at all, which we call the ``unknown-unknowns''.  An oracle that is capable of evaluating the test results of SO algorithm at least has to be able to handle the ``known-unknowns''.

	For an instance of ``known-unknowns'' consider a smart energy grid setting (which is outlined in more detail in \cref{sec:AVPP}) where different power plants are self-organized in different so-called autonomous virtual power plants. 
	If weather-dependent power plant are include, the SO here will depend on the weather conditions. 
	We know that there are different conditions like sunny, rainy, and windy, but we also know that we do not know all different possible combinations and the according correct organizational structures at design-time of the test (or at least we cannot compute all).
	However, an oracle has to cope with that and has to decide whether a result is accepted as correct or rejected as incorrect. 
	
	\item[C-BranchingStateSpace]
	A huge state space is a common challenge for software testing~\cite{Binder1999}, but, as for the oracle problem, SO algorithms add a further dimension. Most of the approaches in software testing coping with a huge state space make use of the structure of the state space to reduce the amount of test cases needed to be executed. For instance, an infinite loop in a code fragment means an infinite state space, but its ramification degree is rather small.  A mechanism applied here is boundary-interior-testing~\cite{Young:2005:STA} that cuts deep branches at certain lengths. SO algorithms, however, are mostly based on heuristics for coping with the ever-changing environment, making the result non-deterministic; these lead to a wide and rather flat branching structure of the state space and make most of the classical techniques hardly applicable directly. 
	
	%\item[C-Isolate] In SOAS, different SO algorithms interact with and influence the system components, possibly resulting in interleaved feedback loops. 
%These are challenging in testing, because it is hard to get dedicated results for a single SO algorithm. 
%To address this challenge, there is a need for isolated testing of single SO algorithms. 
	%\item[C-ErrorMask] The considered SO algorithms provide new system configurations, which could be invalid, and have to be applied to the existing system structure. 
%Under certain circumstances, it is not possible to detect an invalid system configuration if only the resulting system structure is evaluated. 
%Therefore, there is a chance of error masking. 
	%\item[C-BranchingStateSpace] The inherent non-deterministic behavior and the ever-changing environment of SOAS not only lead to a huge state space but also to a tremendous amount of system traces. 
%Especially in SOAS, the latter is problematic since the order of taking particular system states is often crucial for detecting a failure. 
%Due to the non-deterministic behavior of the SO algorithms, there are traces that include a error but do not lead to failure that can be detected by a test oracle (special kind of error masking C-ErrorMask).
\end{description}

We address the challenges for testing SO algorithms in our framework \emph{Isolated Testing of Self-organization Algorithms} ({\framework}) encompassing test case generation, execution, and evaluation. 
It provides techniques and concepts for methodically decomposing and isolating the SOuT and executing them in a controlled environment. 
The isolation can take place on several levels, e.g., isolation from other SO algorithms or from the environment controlled by the SO algorithm itself. 
Addressing C-Isolate further enables to cope with C-ErrorMask since we can reduce disturbances with the testing environment.
However, it turns out that this measure is not enough to address C-ErrorMask, since for SO algorithms operating in several phases the error masking may occur within the SO algorithm itself. We hence introduce a gray-box access to the SOuT by the oracle to be able to detect failures hidden in the depth of the SO algorithms.
This is embedded in the automated oracle of {\framework} whose implementation addresses C-Oracle based on the concept of what we call the corridor of correct behavior.
This corridor allows to distinguish between correct and incorrect states even for the ``know-unknowns'', making it capable for evaluating the test results of the SO algorithm.
Test case generation within {\framework} is based on a static and dynamic test model concept that is capable of handling the complex environment of SO algorithms, addressing C-BranchingStateSpace by systematically selecting test cases based on their occurrence probability within the application domain. 
%For executing the derived test cases a scaffolding concept is important for reproducible test runs and enables the selection of realistic test cases that still adequately reflect the environment, the SO algorithms actually work~in. 

The remainder of this paper is structured as follows. 
\Cref{sec:AVPP} introduces the smart-grid case study used for evaluation and \cref{sec:tesos} presents our main ideas for testing SOAS. 
In \cref{sec:framework}, we detail the building blocks of the {\framework} framework including the test models. 
%\Cref{sec:tesos} introduces our main idea for testing SOAS followed by a short description of the smart-grid case study (cf.\ \cref{sec:AVPP}) used for evaluation purposes. 
%In \cref{sec:framework}, we present the building blocks of the mentioned framework including the presentation of the test model. 
The framework is then used for testing two different SO algorithms introduced in \cref{sec:testedSOalgorithms}.
The evaluation in \cref{sec:evaluation} confirms that our approach encompasses an efficient combination of model-based and random generation techniques based on our test models for finding different kinds of (injected) faults in the examined SO algorithms. \Cref{sec:relatedwork} discusses related work. We conclude with a summary and some ideas for future work in \cref{sec:conclusion}.

\section{Case Study: Self-organized Creation of Virtual Power Plants in Smart Grids}
\label{sec:AVPP}
The wide-spread installation of weather-dependent power plants as well as the advent of new consumer types like electric vehicles put a lot of strain on power grids.
Additionally, small dispatchable power plants (e.g., biogas plants) owned by individuals or cooperatives feed in power without external control.
%Current plans are therefore to scale the controllable output further by installing additional flexible dispatchable power plants and to drive the expansion of the power grid forward.
To save expenses, gain more flexibility, and deal with uncertainties, future \emph{autonomous} power management systems have to take advantage of the full potential of dispatchable prosumers\footnote{We use the term ``prosumer'' to refer to producers as well as consumers.} by incorporating them into the scheduling scheme.
Further, %aleatoric
uncertainties have to be anticipated when creating schedules and compensated for locally to prevent their propagation through the system.

\begin{figure}[!ht]
	\centering
		\includegraphics[width=0.8\textwidth]{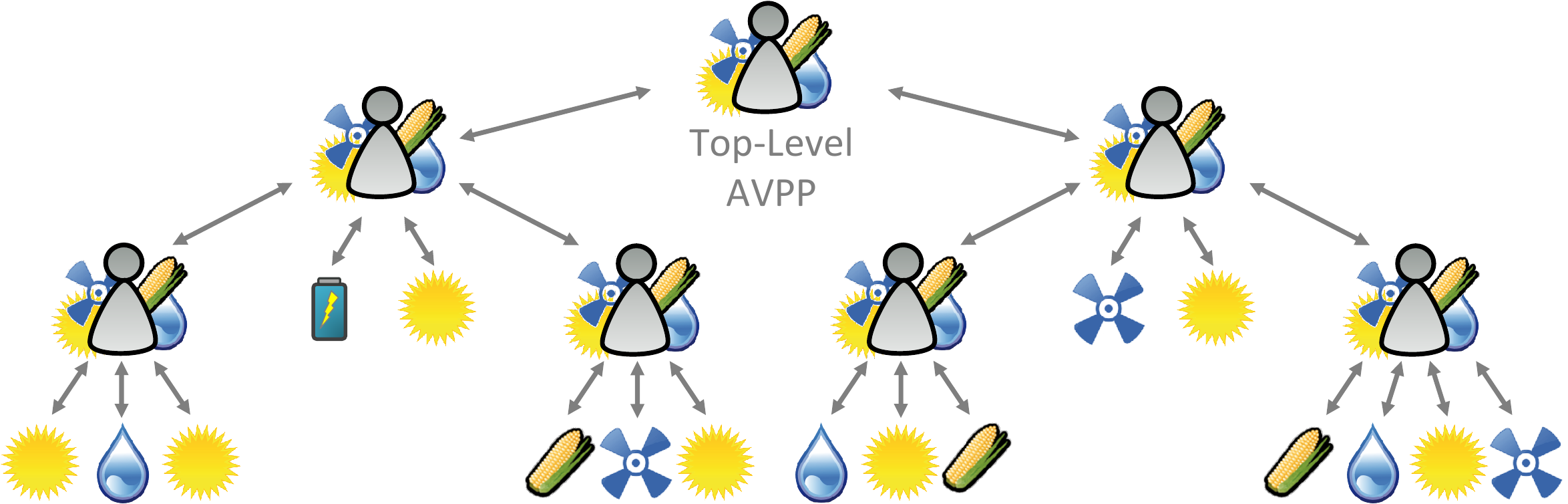}
		\caption{Hierarchical system structure of a future autonomous and decentralized power management system: Power plants are structured into systems of systems represented by AVPPs that act as intermediaries to decrease the complexity of control and scheduling. AVPPs can be part of other~AVPPs.
	The left child of the top-level AVPP, for instance, controls a solar power plant, a storage battery, and two subordinate AVPPs.}
		\label{fig:avpp-hierarchical-system-structure}
\end{figure}

To meet the challenges of future power management systems, Steghöfer et al. presented the concept of \emph{Autonomous Virtual Power Plants}~(AVPPs) in~\cite{steghoefer2013sos} (similar visions of virtual power plants are discussed in \cite{ramchurn2012smart}).
AVPPs represent self-organizing groups of two or more power plants of various types (cf.\ \cref{fig:avpp-hierarchical-system-structure}). The organizational structure represents a \emph{partitioning}, i.e., every power plant is a member of exactly one AVPP, which is established and maintained by a (partitioning-based) SO algorithm. Constraints that specify valid partitionings, e.g., a maximum number of power plants that may belong to one AVPP or that every power plant has to belong to exactly one AVPP, among others, induce a \emph{corridor of correct behavior} over the space of all partitionings.
In this setting, each AVPP has to satisfy a fraction of the overall demand. To accomplish this task, each AVPP autonomously and periodically calculates schedules for directly subordinate dispatchable power plants.
Further, each AVPP's dispatchable power plants have to reactively compensate for deviations resulting from local output or load fluctuations (i.e., uncertainties) to avoid affecting other parts of the system.

AVPPs autonomously adapt their structure to changing internal or environmental conditions, they are able to live up to the responsibility of maintaining an organizational structure enabling the system to hold the balance between energy supply and demand.
In particular, if an AVPP repeatedly cannot satisfy its assigned fraction of the overall demand or compensate for its local uncertainties, %output or load fluctuations locally,
it triggers a reorganization of the partitioning.
The goal is to form homogeneous partitionings in the sense of a structure of similar AVPPs that are likely to feature a heterogeneous composition:
On the one hand, by distributing unreliable power plants among AVPPs, the chance of fluctuations is reduced and the system's robustness increases.
On the other hand, by balancing the AVPPs' degrees of freedom (e.g., by mixing different generator types), their ability to locally deal with uncertainties, i.e., fluctuations, is promoted.
To cope with the vast number of dispatchable power plants, the concept of AVPPs proposes a scalable, hierarchical structure in which AVPPs act as intermediaries. 
This system decomposition reduces the number of dispatchable power plants (including directly subordinate AVPPs) each AVPP controls resulting in shorter scheduling times for each AVPP and the overall system.

In the smart-grid application, we have to cope with C-Isolate as well as C-ErrorMask since the partitioning algorithm and the mechanism that balances supply and demand in each AVPP influence each other significantly. Error masking can also occur in the partitioning algorithms if, e.g., the result of the SO algorithm is faulty but the system state after the reorganization process is valid. The problem of state space explosion (C-BranchingStateSpace) has to be tackled in this case study due to the stochastic nature of the partitioning algorithms in order to deal with the large search spaces, the non-deterministic behavior of single agents, and the stochastic environment (e.g., weather conditions, market prices). 
Finally, C-Oracle occurs in this context since it is hardly decidable at design-time which partitioning for which power plants is correct as this depends on the unknown environmental setting of the controlled power plants as well as on the unknown states of the autonomous power plants themselves.

% As we focus on the self-organized creation of partitionings for the evaluation of our framework, without loss of generality, we only regard a ``flat'' structure of AVPPs in the following sections. In this structure, power plants self-organize into a single layer of AVPPs that resides directly below the root, i.e., the top-level AVPP. In \cref{sec:testedSOalgorithms}, we present the algorithms called SPADA and \psosp\, for the self-organized creation of partitionings.

\section{The Corridor Enforcing Infrastructure (CEI) for Testing Self-organizing, Adaptive Systems}\label{sec:tesos}
\begin{figure}[!ht]
	\centering
	\includegraphics[width=0.8\textwidth]{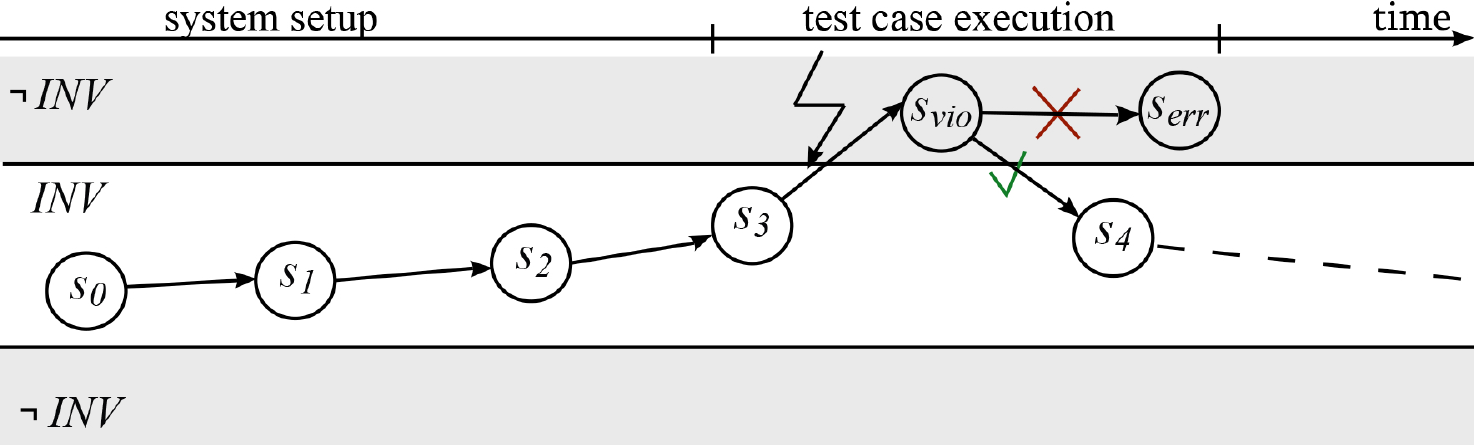}
	\caption{Schematic state-based view of the corridor of correct behavior and the different phases of testing SO algorithms. $\mathit{INV}$ is the conjunction of all constraints of the system controlled by the CEI.}
	\label{fig:ceiScheme}
\end{figure}

Our approach for testing SOAS---and consequently for testing SO algorithms---is based on the \emph{Corridor Enforcing Infrastructure} (CEI)~\cite{Eberhardinger2014TeSOS}. % whose conceptual view is shown in \cref{fig:ceiLevels}.
%As shown in \cref{fig:ceiLevels} we apply a layered testing approach for agent, interaction, and system layer, the framework {\framework} is responsible for testing the controller of the interaction layer (here the SO algorithms are located).
The CEI is an architectural pattern for SOAS using decentralized feedback-loops to monitor and control single agents or small groups of agents in order to ensure that the system's requirements are fulfilled at run-time.
Within the CEI the concepts and fundamentals of the \emph{Restore Invariant Approach} (RIA)~\cite{DBLP:conf/saso/GudemannNOSR08} are applied.
RIA defines the \emph{Corridor of Correct Behavior} (CCB), which is described by requirements concerning the system's structure, formalized as constraints. 
Concerning the smart-grid scenario (introduced in \cref{sec:AVPP}) the CCB is formed by the constraints describing valid  partitionings for the AVPPs, e.g., a maximum and minimum number of power plants that are allowed in each AVPP.
The conjunction of all these constraints is called the \emph{invariant} ($\mathit{INV}$). 
An exemplary corridor is shown in \cref{fig:ceiScheme}. 
If $\mathit{INV}$ is satisfied, the system is inside the corridor; 
otherwise, the system leaves the corridor, indicated by the flash. 
%This requires that the system is transitioned into a safe mode. 
At this point the system has to be reorganized in order to return into the corridor, as shown by the transition with a check mark. 
%RIA states that, if the SO algorithm or a couple of SO algorithms---which are responsible for calculating new system configurations to return into the CCB---find a solution for the current problem and the solution fulfills all constraints and INV holds again. 
A failure occurs in this context if a transition outside of the corridor is taken, indicated by a cross.

\begin{figure}[!t]
	\centering
	\includegraphics[width=0.85\textwidth]{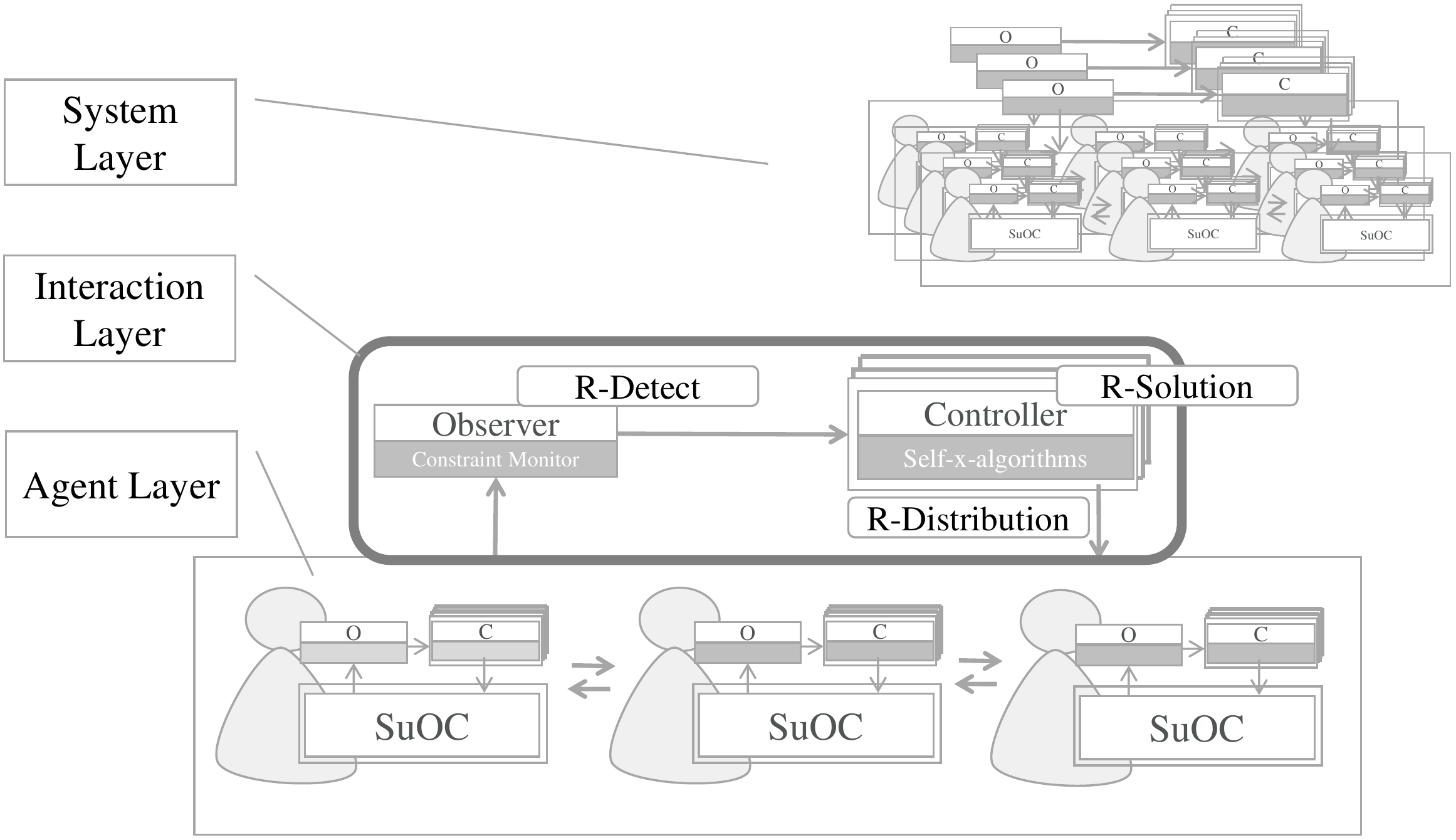}
	\caption{Schematic view of the \emph{Corridor Enforcing Infrastructure} (CEI) and its different level of testing (namely, agent, interaction, and system layer). The {\framework} Framework is located in this concept on the \emph{Interaction Layer} where the SO algorithms are incorporated into the \emph{Controller}. Besides the \emph{Controller (C)} the CEI main components are the \emph{System under Observation and Control (SuOC)}, and the \emph{Controller (C)} that form together distributed, decentralized feedback-loops on different levels, i.e., the CEI consists sets of nested feedback-loops (symbolized by the nested set of different system parts in the upper part of the figure) controlling the entire system.}
	\label{fig:ceiLevels}
\end{figure}

The CEI implements the RIA with decentralized pairs of a monitor and a controller, similar to the MAPE cycle~\cite{kephart2003vision} or the Observer/Controller (O/C) architecture~\cite{Schmeck2010}.
The schematic view in \cref{fig:ceiLevels} shows an implementation of the CEI based on the O/C architecture where the essential parts are the system under observation and control~(SuOC, i.e., single agents or groups of agents controlled), the observer (O, i.e., the component monitoring the state of the SuOC and providing information to the controller) and the controller (C, i.e., the self-x-algorithms controlling the SuOC, e.g., an SO algorithm).
Note that the CEI consists of sets of nested feedback-loops controlling the entire system.
\Cref{fig:ceiLevels} shows further the different layers of testing applied to cope with the complexity of the system: agent, interaction, and system layer.
The {\framework} framework is located in this concept on the \emph{Interaction Layer} where the SO algorithms are incorporated into the \emph{Controller}.

In some cases the observer for single agents or small groups of agents can be generated (semi-)automatically from the requirements documents for the considered system, as shown by Eberhardinger et al.~\cite{Eberhardinger2013}.
The small groups controlled in the smart-grid scenario are the power plants partitioned into AVPPS.  Furthermore, single power plants are in control loops to adjust, in particular, their energy production.
The reorganization by the controller is performed by one or more SO algorithms resulting in a new system configuration.
Such a system configuration has to satisfy the constraints describing valid organizational structures, e.g., a maximum number of controlled components within each organization.
With regard to the partitioning problem---applied for the AVPPs---this means that each power plant belongs to exactly one AVPP.
For other application scenarios one might require a structure in which each component belongs to at least one organization, which corresponds to a set covering~\cite{Balas1976SetPartitioning}.
The concrete choice of the organization algorithms and their constraints has no impact on our approach, both set covering and partitioning SO algorithms could be implemented within the concept of the CEI.
%Inside the CCB, the system behaves like a traditional software system and traditional test techniques can be used to ensure the quality of the \emph{System under Observation and Control} (SuOC). 
%The CEI instead is responsible for the self-organizing and adaptive behavior of SOAS. 

In order to grasp SO algorithms for testing, we need techniques to examine the CEI and its mechanisms, covering the following responsibilities of the CEI (marked with a rounded rectangle in \cref{fig:ceiLevels}): correct initiation of a reorganization if and only if a constraint is violated (monitoring infrastructure, R-Detect); calculation of correct system configurations in case of violations (R-Solution); and correct distribution of these configurations within the single agents or small groups of agents controlled by the CEI (R-Distribution). 
% In fact, this can be done separately from the rest of the system which consequently reduces the test effort. 
% In fact, several different failures can occur due to the SO mechanisms, which are part of the controller (cf.\ \cref{fig:ceiLevels}).  
At the moment, {\framework} focuses on revealing the kinds of SO algorithms failures related to (R-Solution) and (R-Distribution). 
\Cref{sec:conclusion} gives an outlook on the possible extensions of the framework for testing the monitoring infrastructure (R-Detect). 

On the other hand, the CCB immediately affords for our testing efforts a check whether a system configuration produced by an SO algorithm for reorganization is correct, i.e., whether all constraints of the CCB hold.
This integrated check consequently forms the basis of a test oracle in the {\framework} framework.
However, in order to use the CCB correctness check for
result validation, a consistent snapshot of the system state is needed. 
Therefore, in a first step, {\framework} is built upon a stepwise execution model that naturally allows to synchronize all system components at distinct points in time.  The necessary synchronization may reduce the possible interleavings of component behaviors, but this may be mitigated by letting components decide for making an idle local step in every synchronous round.
Still, the concept of the CCB makes it possible to create a fully automated oracle that allows to tackle C-Oracle, since it is possible to decide even under unknown conditions of execution whether a state is inside the CCB or not, addressing an evaluation of the ``known-unknowns''.

\section{A Framework for Isolated Testing of Self-organization Algorithms (\framework)}\label{sec:framework}
\begin{figure}[!t]
	\centering
		\includegraphics[width=0.95\textwidth]{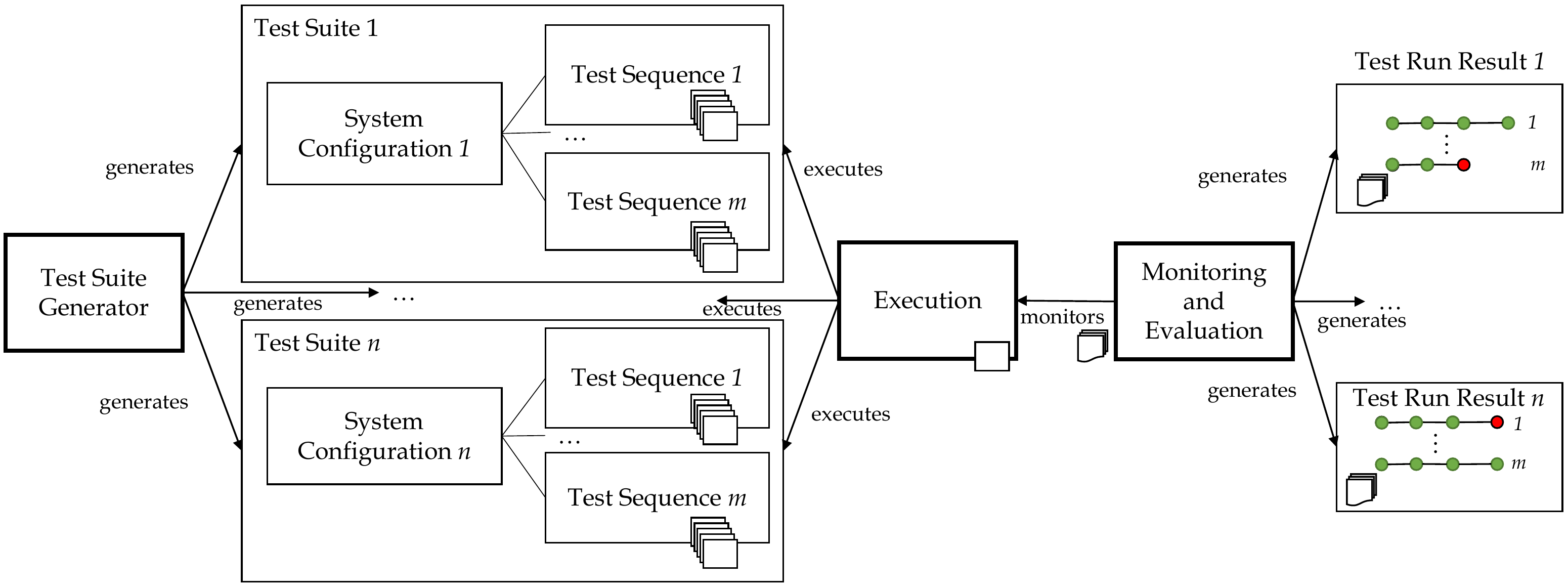}
	\caption{{\framework}  generates $n$ test suites, each consisting of an initial system configuration for the test setup and $m$ test sequences. Each test sequence contains a number of test cases (indicated by the overlapping rectangles). Each test sequence is executed individually (indicated by the single rectangle in the execution component). The monitoring and evaluation component monitors the system and samples the reorganization results as well as the system structure for each test case (indicated by the forms at the arrow between the monitoring and evaluation component and the execution component). For each test suite, the results of all test cases are evaluated in a test run~result, leading to $n$ results for $m$ evaluated test sequences.}
	\label{fig:overviewFramework}
\end{figure}

As common for testing frameworks, we organize {\framework} in a process-oriented manner into: test case generation\footnote{Due to the fully automated evaluation of test cases by the oracle component of {\framework}, test case generation  reduces to test input generation as no expected output is needed. This concept builds up on Artho et al.~\cite{artho2005combining} also combining run-time verification and test input generation for creating test cases. In the remainder of this paper we use test case generation in the sense of test input generation.} (cf.\ \cref{sec:inputGeneration}), test execution (cf.\ \cref{subsec:executionComponent}), and, finally, test evaluation as a part of output processing (cf.\ \cref{sec:outputComponent}). An overview of the framework is shown in \cref{fig:overviewFramework}.

To derive appropriate test suites within the test suite generator component, we use the test model presented in \cref{sec:testModel} that enables sufficient abstraction to tackle C-BranchingStateSpace.
{\framework} allows for isolating SO algorithms from other parts of the system, including other SO algorithms (C-Isolate). 
For this purpose, {\framework} provides a fully controlled test environment, i.e., the environment of the SOuT is fully mocked. 
To enable mocking, the execution component, described in \cref{subsec:executionComponent}, provides a system simulator and a gray-box interface for the SOuT. The gray-box interface enables the evaluation of the internal state of the SOuT, e.g., for checking interim results, and thus also allows us to address C-ErrorMask. 
%The information needed by the oracle to detect these overlapping failures are provided in the form of interim results by the SOuT.
The evaluation of the validity of reconfigured system structures is performed by the test oracle that checks the violation of constraints describing valid solutions (C-Oracle). The responsible component for the test evaluation---the monitoring and evaluation component---is described in \cref{subsec:evalcComponent}.
 
Overall, we assume that the test system is embedded into an agent-based execution system. To allow a consistent snapshot and therefore coherent synchronization points after each executed test case, the system is based on a stepwise execution model. 

\subsection{Test Model of the Framework {\framework}}\label{sec:testModel}
The test model of the framework is the source for generating test suites (cf.\ \cref{sec:inputGeneration}) and is composed of three parts:
\begin{enumerate}[leftmargin=*]
	\item Model of the system under test (SuT)
	\item Model of the SO algorithm under test (SOuT)
	\item Model of environmental changes and influences 
\end{enumerate}
The first one provides the necessary information for the domain in which the SOuT is applied, the second one defines suitable configurations for the SOuT which are highly influenced by domain knowledge. The model of environmental changes and influences describes the dynamics of the environment and is mainly used for creating test sequences. In the following, these three parts of the test model are described with respect to their intended purpose within the test suite generation component.

\paragraph{Model of the system under test.}\label{subsec:staticTestModel}
%For generating the initial system configuration of a test run, a model of the test setting is needed. 
%The intention is to provide the information necessary to automatically generate and also reasonably constraint the following setting within a test run (as described in \cref{sec:inputGeneration}):
%\begin{itemize}
	%\item Set of agents of different types
	%\item Initial state of the agents
	%\item Agent groups and members
	%\item Initial system structure
	%\item Fully parametrized SO algorithm under test (SOuT)
%\end{itemize}
%The needed specification is further subdivided into the test system and the SOuT.
%\paragraph{Modeling the Test System}
The model of the SuT specifies all important information from the domain the SOuT belongs to and is used for generating the system configuration within the test suites. 
For this purpose, a domain description (in the form of a UML class diagram) is provided that is enriched by constraints describing the CCB (cf.\ \cref{sec:tesos}). 
With regard to our case study, the constraints define valid partitionings, e.g., the maximum and minimum number of power plants for each AVPP as well as the constraint that each power plant must be contained in exactly one AVPP.
Further, the behavior of each agent type (e.g., wind turbines, solar panels, or biogas power plants) is defined by standard design documents. 
In our smart-grid case study we additionally have to define \emph{groups of agents} which are influenced by similar environmental changes, such as wind turbines with a certain locality.
Accordingly, a group of agents share one \emph{environment profile}, a stochastic model abstracting possible environmental changes (more details are given at the end of this subsection). 
The fact that an agent is in a certain group, is not known to the SOuT and has no direct influence on the partitioning decisions of the SOuT. 
The members of a group of agents can be of different agent types.

We define how the environment influences certain types of agents so that a reduction of the test cases to realistic scenarios is possible.  
Indeed, describing the influences relation between the environment and types of agents is rather complex.
At this step we rely on simplification and abstraction within the model to keep our approach scalable: We assume that the mapping of environment influences to agent types (in its abstracted form) can be determined and defined a priori, i.e., we neglect that the influence might change over time or is indeterministic.
%Based on this assumption a mapping function could be defined between the environment state (relevant for the concerning group) and states of the agents within the group.
%Thus, the influence might be described by mapping of delta-values, describing the change of the current state according to the appearance of an environment state.
The consequences of this assumption and simplification are on the one hand that the model is scalable within the approach and can be handled by the test engineer, but, on the other hand, that it might neglect some situations for testing.
Our evaluation results (cf.\ \cref{sec:eval-results}), however, showed that it is still possible to find different kind of failures. 

%The information which is described in this model encompass the specification used for generating the system configurations of the test runs (cf. \cref{fig:overviewFramework}). 
%For this purpose, a domain description (in form of an \emph{UML class diagram}) is used that is enriched by constraints for codomains for the classes and their states. 
%The minimal subset which have to be described in the model is:
%\newpage
To recap, the model of the system under test must contain at least:
\begin{itemize}[leftmargin=*]
	\item Types of agents
	\item Definition of possible initial states for the agents
	\item Suitable ranges for the minimum and maximum number of agents of a specific agent type
	\item Constraints concerning groups of agents (minimum and maximum number as well as size)
	\item Mapping of environmental influences to agent types
	\item Constraints concerning valid system structures, i.e., the relevant part of the CCB
\end{itemize}
%
%Additional parameters for the test cases generation are a minimum and maximum number of test cases within each generated test sequence and the number of test sequences that should be created for each test suite.

Since the model of the system under test depends on the application domain the description for the model itself is quite generic and coarse.
However, the detailed information of the application---given in \cref{sec:evaluation}---shows how to transfer this generic description into a specific model of the system under test.

\paragraph{Model of the SO algorithm under test.}
This model specifies valid configurations for the SOuT. 
It is used to derive valid ranges for all relevant parameters, such as the algorithm's maximum run-time. 
The selection of relevant parameters highly depends on the concrete algorithm (e.g., some algorithms allow to specify a maximum execution time and some do not), the situations that should be covered by the test runs (e.g., some failures only occur if we give the algorithm enough time), and the domain knowledge (e.g., in some domains, the maximum run-time is naturally bounded).
Clearly, despite the random testing approach, a suitable parametrization can be used for directed testing. 
This means that we can push the algorithm into interesting  directions, e.g., to use specific functionality, which increases the code coverage.  

The parameters of the SOuT and dependencies between the parameters are specified as constraints of the permitted setting of the parameters of the SOuT. 
%The model is specified with variables for the parameter that are restricted by domains and relations between them.
The resulting constraint satisfaction problem (CSP) is solved by the input component of the framework for forming a system configuration within a test run. 
For a SOuT that is based on a particle swarm optimization algorithm to form organizational structures (like the PSOPP algorithm described in \cref{sec:pso}) a parameter of the algorithm is the number of particles to use, constrained in the model of the SOuT, for instance, by an upper and lower bound.
Further, the number of particles are related to the constraints concerning the number of partitions to be formed.
This relationship is incorporated into the CSP. 
The described CSP is the foundation for automatically generating valid settings for the SOuT.
A more detailed example for the specification is given in \cref{sec:evaluation}. 
%The second aspect in the description of the test setting is the model of the SOuT.

\paragraph{Model of environmental changes and influences.}\label{sec:ep}
Recalling our idea of isolated testing of SO algorithms, the third component of the test model is the model of environmental changes and influences.
This is because of possible interferences with other SO algorithms due to interleaved feedback loops (C-Isolate) as well as the huge state space induced by the different possible states of the environment and the algorithm's non-deterministic behavior (C-BranchingStateSpace).
We address C-BranchingStateSpace by providing stochastic models of the environment, called \emph{environment profiles}~(EPs), and C-Isolate by \emph{functions describing the environment's influence} on the system that enables to decouple the SO algorithm.\footnote{Note that the environment also covers in this case other SO algorithms of the system.}

As we do not assume that all agents share the same environment (e.g., because of their geographical distribution) and are equally influenced by environmental conditions, we define these EPs and influence functions with regard to a specific group of agents~$\mathcal{G}$.\footnote{That technique of state reduction is performed according to the state abstraction principles that are well known in classical testing~\cite{Young:2005:STA}.}
This allows us to deal with large state spaces even better (C-BranchingStateSpace).
For example, it is not necessary to consider the complete set of possible states of the environment $\mathcal{E} = \{\mathsf{(cloudy, high\ price)},$ $\mathsf{(rainy, high\ price)}$, $\mathsf{(sunny, high\ price)}$, $\mathsf{(cloudy, low\ price)}$, $\mathsf{(rainy, low\ price)}$, $\mathsf{(sunny,}$ $\mathsf{low\ price)}\}$ if we regard a group of solar power plants whose output mainly depends on the current weather conditions and is more or less independent of the current market price (a property that is defined in the mapping of environmental influences to agent types).
Instead, for each group of agents~$\mathcal{G}$, we map one or more states of the environment from the set $\mathcal{E}$ to a single so-called \emph{relevant state} that describes the relevant parts of the environment's state for $\mathcal{G}$, i.e., those that have an influence on $\mathcal{G}$'s behavior. By gathering all relevant states, we obtain the entire set of relevant states $\mathcal{R}_{\mathcal{G}}$ for $\mathcal{G}$.

In case of our group of solar power plants the states $\mathsf{(cloudy,}$ $\mathsf{high\ price)},$ $\mathsf{(cloudy, low\ price)} \in$ $\mathcal{E}$ are mapped to a state $\mathsf{cloudy}$ that becomes a member of $\mathcal{R}_{\mathcal{G}}$. In this example, $\mathcal{R}_{\mathcal{G}}$ is finally equivalent to the set of weather conditions $\{\mathsf{cloudy},$ $\mathsf{rainy},$ $\mathsf{sunny}\}$ considered in $\mathcal{E}$.

The identification of relevant states is supported by the mapping of environmental influences to agent types as described in the model of the system under test.
Thus, if the environment state or a set of environment states corresponds to an environmental influence that is already mapped to an agent type of the concerning agent group, this state has to be included into the set of relevant states for this agent group.
However, mapping the environment's states to relevant states of an agent group is---as the mapping of environment influences to agent types in the model of the system under test---not generically solvable; in every application domain this classification of relevance has to be made specifically.
Further, the relevant states are not limited to the mapping described for the environmental influences to agent types, since, among other things, the other SO algorithms are here also part of the environment (where as that is not considered by the model of the system under test).

The exemplified description above shows what the necessary steps are and how this mapping should be achieved. 
The mapping in general does not have to be disjoint but we expect it to be complete since only useful environment states should be included in $\mathcal{E}$, i.e., states that influence at least one of the agent groups.

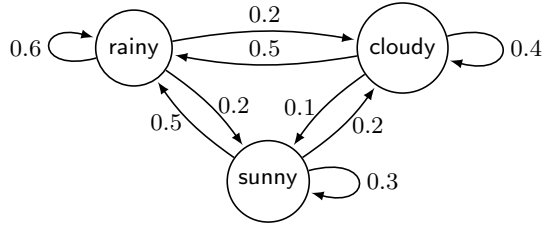
\begin{figure}[t]
\centering
\begin{tikzpicture}[->,>=stealth',shorten >=1pt,auto,node distance=2.5cm,
semithick]
\tikzstyle{every state}=[fill=none,draw=black,text=black]
\node[state] (sunny) {$\mathsf{sunny}$};
\node[state] (cloudy) [above right of=sunny] {$\mathsf{cloudy}$};
\node[state] (rainy) [above left of=sunny] {$\mathsf{rainy}$};
\draw[-latex] (sunny) to[bend right=10] node [right] {$0.2$} (cloudy);
\draw[-latex] (sunny) to[bend left=10] node [left] {$0.5$} (rainy);
\draw[-latex] (sunny) to[loop right] node[right] {$0.3$} (sunny);
\draw[-latex] (rainy) to[bend left=10] node [above] {$0.2$} (cloudy);
\draw[-latex] (rainy) to[bend left=10] node [right] {$0.2$} (sunny);
\draw[-latex] (rainy) to[loop left] node[left] {$0.6$} (rainy);
\draw[-latex] (cloudy) to[bend left=10] node [above] {$0.5$} (rainy);
\draw[-latex] (cloudy) to[bend right=10] node [left] {$0.1$} (sunny);
\draw[-latex] (cloudy) to[loop right] node[right] {$0.4$} (cloudy);
\end{tikzpicture}
\caption{A simplified EP for a group of solar power plants at a specific geographic location: Possible weather changes between $\mathsf{rainy}$, $\mathsf{sunny}$, and $\mathsf{cloudy}$ are indicated by directed edges. The numbers represent transition probabilities.}
\label{fig:exampleEPweather}
\end{figure}

With regard to a specific group of agents, an EP not only captures the relevant states $\mathcal{R}_{\mathcal{G}}$ of $\mathcal{G}$'s environment, but also probabilities for changes from one state to another.
Assuming that the next state only depends on the current state, an EP represents a first-order Markov chain.
\cref{fig:exampleEPweather} depicts a simplified example of an EP for a group of solar power plants in a specific region; as a matter of fact, the models used for testing are much more complex (cf.\ \cref{sec:evaluation}).% and this example here is only used for illustration purposes.
Such an EP can either be created using domain knowledge, derived from statistical data gathered during the execution of the system under test, or a combination of both.
In the literature of multi-agent systems, Markov chains are often used to simulate the environment for evaluation purposes (cf.\ \cite{scott2013Residential,anders2014proactive}). 

To model the way the environment influences the members of an agent group~$\mathcal{G}$, we use a function $f_\mathcal{G}: \mathcal{R}_{\mathcal{G}} \times \mathcal{S}_\mathcal{G} \rightarrow \mathcal{S}_\mathcal{G}$, where $\mathcal{S}_\mathcal{G}$ represents all possible states of $\mathcal{G}$'s members.
With regard to a member $a \in \mathcal{G}$, the function $f_\mathcal{G}$ maps the new state~$\sigma_{\mathit{env}}^\prime \in \mathcal{R}_{\mathcal{G}}$ of $\mathcal{G}$'s environment and $a$'s current state~$\sigma_a \in \mathcal{S}_\mathcal{G}$ to a new state~$\sigma_a^\prime \in \mathcal{S}_\mathcal{G}$.
For instance, the change of the current weather conditions from $\mathsf{sunny}$ to $\sigma_\mathit{env}^\prime = \mathsf{rainy}$ could impair a solar power plant's ability to make adequate predictions of its future output, which is reflected in the transition from $\sigma_a = \mathsf{good\ predictions}$ to $\sigma_a^\prime = \mathsf{bad\ predictions}$.
Different influences of the weather on different types of weather-dependent power plants---represented as different agent groups---can be formalized by group-specific functions~$f_\mathcal{G}$.
On the other hand, if an EP describes possible developments of the prices at an energy market, $f_\mathcal{G}$ can model the way a power plant or consumer behaves at the market, i.e., its strategy.
For example, if the market price falls below a certain threshold, some consumers might change their strategy to ``buy energy'', whereas the producers might become more reluctant to sell their production.
As is the case with the creation of EPs, such influence functions can be deduced from domain knowledge and statistical data.
Note that, in some cases, the influence function might depend on random variables and thus becomes probabilistic.
This reflects the fact that we cannot assume perfect knowledge about the influences of the environment on the agents due to the high complexity and the agents' autonomy.
However, it is depends on the test designer's choice as well as on the application domain whether to use a probabilistic functions to map the environment influences to the states of agent groups.
The drawback of modeling these functions on random variables is that the process of test case generation is less controllable, but it might reveal some interesting test cases.
In our evaluation (cf.\ \cref{sec:evaluation}), we used deterministic functions which especially payed off in the process of getting more control over the test case generation procedure and gaining higher failure detection rates. 
%This reflects the overall goal of testing that is finding failure, thus it is worthy to abstract from some characteristics if that leads to better handling in test case generation procedure in order to gain a higher failure detection rate.
Of course, there might be applications where detailing the model applies better in order to gain a higher failure detection rate due to the more accurate model that is for instance able to reveal border cases that are failure prone.
The choice depends here on the test designer that adapts the presented approach to specific algorithms and application domains.  

As we will explain in the next subsection, the EPs are used for the test case generation by simulating the Markov chains, yielding random but representative test sequences describing environmental changes for the different groups of agents.
The probabilistic information of the EPs can be used to estimate the relevance of generated test cases and test sequences and their likelihood of occurrence in a realistic setting (C-BranchingStateSpace).
As we will see in our evaluation in \cref{sec:evaluation}, this property makes EPs also an important tool for coverage analysis.

\subsection{Test Suite Generator Component of \framework}\label{sec:inputGeneration}
{\framework}'s test suite generator component generates different test suites each containing \emph{one} initial system configuration on which basis it randomly creates $\mathit{n}$ test sequences by using the environment profiles (cf.\ \cref{fig:overviewFramework}).% and, for more details,~\cref{fig:Component_Diagram__Testsystem}). 
A randomly generated initial system configuration contains the following information:
\begin{itemize}[leftmargin=*]
	\item a set of agents, possibly of different types
	\item a set of initial agent states, one for each agent
	\item a set of agent groups so that each agent is contained in exactly one group
	\item an initial system structure
	\item the selected SOuT
	\item additional system parameters, including the parametrization of SOuT and a random seed to be used during the execution of the test sequences
\end{itemize}
The fact that all sequences within one test suite start with the same system configuration is important in the context of SOAS to tackle the already mentioned four challenges (cf.\ \cref{sec:intro}). Each test sequence is an ordered series of test cases whose length corresponds to the number of discrete (time) steps that should be simulated in the course of the test run. 
This is possible since {\framework} uses a stepwise execution model. To generate the test cases, a chosen set of EPs is simulated for the specified number of time steps to obtain the sequence of environmental changes. 

{\framework}'s test suite generator component can be used in two different modes concerning the execution of the framework: 
\begin{itemize}[leftmargin=*]
	\item \emph{Online: } This mode produces new test suites during the execution of the framework. It accordingly enables an ``endless'' execution of the framework. For this purpose a new test suite is generated and executed after every completed test run.
	\item \emph{Offline: } This mode assumes a given set of test suites which is generated a priori and sequentially executed by the framwork.  
\end{itemize}
In both modes, the test suites are executed by the execution component.

\subsection{Execution Component of \framework}\label{subsec:executionComponent}

The execution component defines two phases: (1)~setting up the system and (2)~executing the test cases on the SOuT.
%This division is based on our general theory of testing SO systems (cf.\ \cref{sec:tesos}) in which a process is split into a setup and a testing phase. 
%The testing phase is then monitored by the output component.

\begin{figure}[t]
	\centering
		\includegraphics[width=1.00\textwidth]{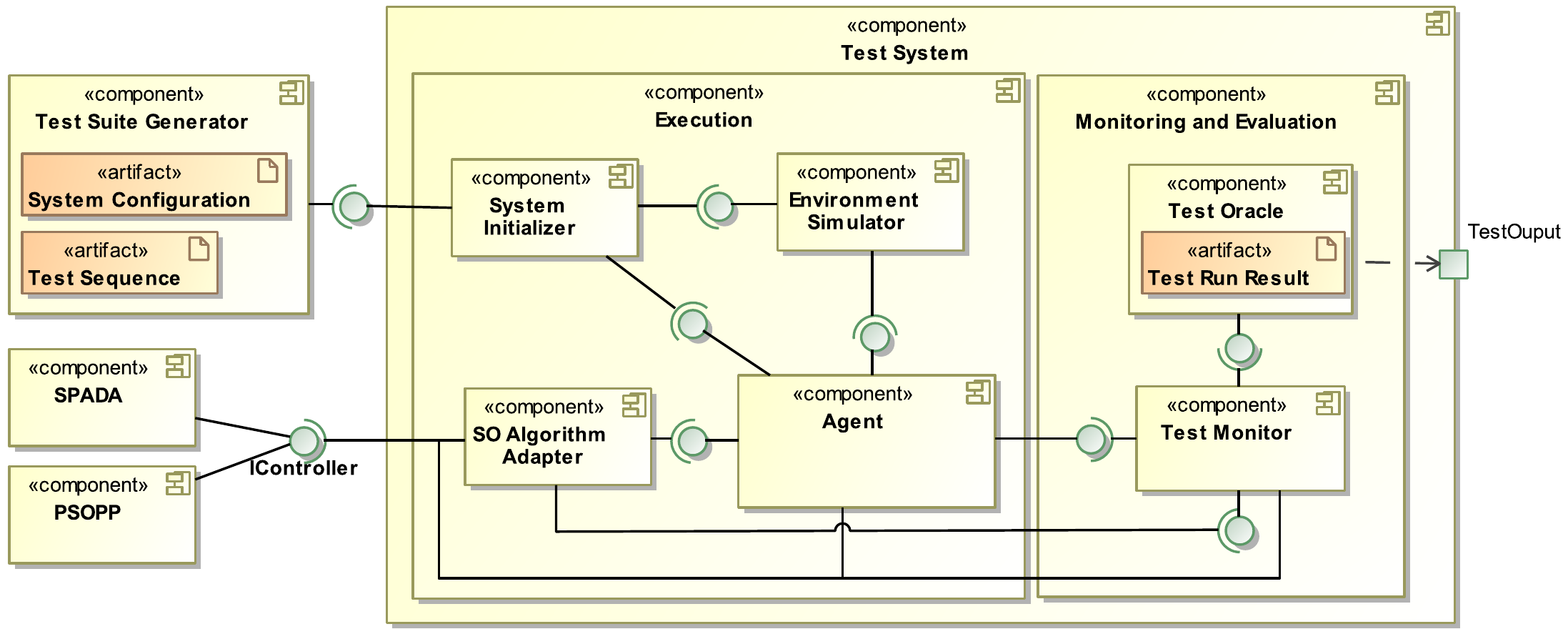}
	\caption{UML component diagram of the detailed architecture of {\framework} having two different partitioning-based SO algorithms plugged in (SPADA, PSOPP). The \emph{Test Suite Generator} component provides the artifacts of a test suite. The \emph{System Initializer} component initializes the system using the generated test suites and sets up the \emph{Environment Simulator} and the \emph{Agents}. Within the \emph{Execution} component, the test suites are executed via the \emph{Environment Simulator} while the \emph{Monitoring and Evaluation} component logs the execution (\emph{Test Monitor}) and the \emph{Test Oracle} evaluates it to \emph{Test Run Results}.}
	\label{fig:Component_Diagram__Testsystem}
\end{figure}

%\begin{sidewaysfigure}
	%\centering
		%\includegraphics[width=1.00\textwidth]{Sequence_Diagram__Testsystem__Testsystem.pdf}
	%\caption{UML sequence diagram viewing the procedure executed of an application of {\framework} for testing SO algorithms. Overall, all necessary steps are shown to execute and evaluate a test run which is provided by the \emph{Test Suite Generator}. Steps 1 to 5 concern the setup of the system while 6 to 15 subsume the execution of test sequences and their evaluation. }
	%\label{fig:Sequence_Diagram__Testsystem__Testsystem}
%\end{sidewaysfigure} 

\paragraph{Setting up the system.}
%To get the system in position to execute the generated test cases in the context of the selected SOuT, several steps are necessary.
First, the \emph{System Initializer} sub-component processes the initial system configuration provided by the current test suite. For this purpose, it sets up the set of predefined agents, using the given initial agent states, and establishes the given initial system structure.
Next the SOuT is incorporated via an interface (\texttt{IController}) which plugs in the SOuT via an adapter component. At this stage, we can use {\framework} only for black-box testing, since we have no access to interim results of the SOuT. For the isolated testing of our partitioning algorithms, we plugged in SPADA and PSO as shown in \cref{fig:Component_Diagram__Testsystem}. The adapter component is an implementation of the adapter pattern that enables efficient testing of different SO algorithms. Therefore, the adapter fulfills the following tasks, when reorganizing the system structure: (i) it instantiates the specified SOuT, (ii) sends the current system structure to this SOuT, (iii) as soon as the SOuT terminates, the adapter informs the test monitor (as described in \cref{sec:outputComponent}) about the solution, and (iv) after the monitor checked the solution, the adapter requests the SOuT to adopt the new system structure. Point (iii) yields a gray-box view for testing, since we are now able to check interim steps of the test case execution. This is needed for testing if the SOuT---as part of the controller of the CEI---fulfills (R-Solution) besides (R-Distribution). 
%\\ \\ 
%Furthermore, for integrating the SOuT in the test system it has to comply to the interface of the test system.
%This interface mainly describes the necessary access specification that are used by in the gray-box approach.
%To these belong among others that the algorithm is following the steps of analyzing the problem, finding a solution and deploying it. 
%The found solution has to be able to be retrieved by the test monitor before deploying. 
%Beside this the further interaction described in this section is assumed to be fulfilled via this interface. 

%\benedikt{At the moment the results are not reproducible, because of among others Java HashMap behavior and process scheduling. How are we going to state that? Is the future work section enough? I don't think so, because the question how we deal with this common issues arises here.}

\paragraph{Executing test cases on the SOuT.}
After the setup is completed, the environment simulator sequentially executes the given test cases. 
This execution is based on a stepwise execution model that enable consistent synchronization points after every test case.
First, the agent components perceive the corresponding environmental changes and adapt their internal states according to their influence functions $f_\mathcal{G}$. 
After the changes are applied to every component, the system constraints are checked for a violation, i.e., whether the system leaves the CCB, if so the SOuT is activated in next step via the adapter component.\footnote{Note that not every test case execution leads directly to constraint violations and thus to an activation of the SOuT. To form a realistic system structure within the test system, it is necessary to allow the system to take transitions that do not violate the CCB.} 
Based on the current system state, the SOuT has to reorganize the system structure in order to restore the system constraints, i.e., to bring the system back into the CCB.
That last action completes one execution sequence that correspond to the executed test case.
%This reorganization is performed in two steps (where the former is observable because of the gray-box approach): 
%\begin{enumerate}[(1)]
	%\item the SOuT determines a new system structure that adheres to the system constraints and
	%\item the computed solution is adopted.
%\end{enumerate}
%{\framework}'s monitoring and evaluation component monitors both steps and evaluates the validity of the computed results as well as the resulting system structures.

%As mentioned before, our test vision is based on the CEI. 
The gray-box view on the SOuT allows us to tackle (R-Solution) as well as C-Oracle since we can evaluate the calculated system structure before it is distributed among the agents. 
In addition, the gray-box view consequently avoids error masking during testing---defined as challenge~C-ErrorMask. 
The adapter continuously informs the test monitor about the current system state during the test case execution at specific points in time.
Gaining the synchronized system state is possible due to the stepwise execution model. 
Based on this information, the test oracle can decide if the distribution of the solution leads to a correct system configuration, thereby addressing (R-Distribution). 

\subsection{Monitoring and Evaluation Component of {\framework}}\label{sec:outputComponent}
The monitoring and evaluation component of {\framework}---that mainly tackles \mbox{C-Oracle} by applying the concepts of the CCB to monitoring and evaluation---is split into two sub-components, each being responsible for a specific task: 
\begin{enumerate}[leftmargin=*]
	\item the test monitor observes and samples data of the executed SOuT and of the test system
	\item the test oracle evaluates the sampled data
\end{enumerate}
To live up to these responsibilities, the monitoring and evaluation component directly interacts with the execution component and the SOuT using the provided interfaces (cf.\ \cref{fig:Component_Diagram__Testsystem}). 

\paragraph{Monitoring \& sampling data.}
In the course of the execution of a test suite, the monitoring and evaluation component monitors and samples data of the test system influenced by the SOuT as well as directly of the SOuT. This allows to observe the two steps of a SOuT: (1) computing a solution for the reorganization, and (2) distributing the solution to the agents. The result of these steps is stored in the current system state, and sent to the test oracle for further evaluation. 
A necessary prerequisite for monitoring and sampling data is the established stepwise execution model, allowing to synchronize all system components at distinct points in time for later evaluation.
However, the synchronization of all system components does not imply prerequisites or restrictions to the SOuT's characteristic.
Having a synchronization point and a global view within the test system does not interfere the decentralization of the SOuT since the SOuT itself makes no use that global view, it is only introduced for test oracle.
Indeed, there are some restrictions concerning the possible tested situations. 
The stepwise execution model neglects test cases that address situations where components are in different states.
Incorporating them into our test approach is a matter of future work outlined in \cref{sec:conclusion}.
%In the upper part of \cref{fig:state_view_evalTestcase}, we show an exemplary state view of three monitored consecutive tests of a test sequence.

%\begin{figure}[t]
	%\centering
	%\begin{tikzpicture}[->,>=stealth',shorten >=1pt,auto,node distance=1cm,
  %thick,main node/.style={circle,fill=white,draw},fault node/.style={circle,fill=red!50,draw},correct node/.style={circle,fill=green!50,draw}]
		%
		%\node[main node] (s0)	{$\tau_0$};
		%\node[main node] (s1)	[right=of s0] {$\tau_1$};
		%\node[main node] (s2) [right=of s1] {$\tau_2$};
		%
		%\node[correct node] (s0_1) [below=of s0]	{$\tau_0$};
		%\node[correct node] (s1_1)	[right=of s0_1] {$\tau_1$};
		%\node[fault node] (s2_1) [right=of s1_1] {$\tau_2$};
		%
	  %\path[every node/.style={font=\sffamily\small}]
			%(s0) edge node [right] {} (s1)
			%(s1) edge node [right] {} (s2)
			%(s0_1) edge node [right] {} (s1_1)
			%(s1_1) edge node [right] {} (s2_1);
		%\path[every node/.style={font=\sffamily\small},dashed]	
			%(s0) edge node [above] [right]{evaluate} (s0_1)
			%(s1) edge node [above] [right]{evaluate} (s1_1)
			%(s2) edge node [above] [right]{evaluate} (s2_1);
		%\begin{scope}[on background layer]
			%\node [fill=black!30,fit=(s0) (s1) (s2)] {};
		%\end{scope}
		%\begin{scope}[on background layer]
			%\node [fill=gray!30,fit=(s0_1) (s1_1) (s2_1)] {};
		%\end{scope}		
	%\end{tikzpicture}
		%\caption{Upper box: sampled states of the tests $\tau_0, ..., \tau_2$ of a specific test sequence. Lower box: each state is evaluated and either marked as \emph{failed} (red) or \emph{passed} (green).}
	%\label{fig:state_view_evalTestcase}
%\end{figure}

\paragraph{Evaluating the data by a test oracle.}\label{subsec:evalcComponent}
Having received the sampled state for a test case from the test monitor, the constraint-based test oracle evaluates the result of the test case either to \emph{passed} or \emph{failed}.
%The evaluation step is shown in \cref{fig:state_view_evalTestcase} by the dashed lines between sampled states of test cases $\tau_0, ..., \tau_2$ within a test sequence $\mathfrak{t}$  which are individually evaluated to either \emph{failed} (indicated by colouring red) or \emph{passed} (indicated by colouring green). 
%This evaluation is performed by the constraint-based test oracle of the framework, which performs the evaluation on the one hand based on the agents as well as their structure within the test system and on the other hand based on the computed solutions of the SOuT.
The evaluation of the results is based on constraints, which represent the requirements for the SOAS as well as the SOuT and form the CCB (as described in \cref{sec:tesos}). Each constraint must be satisfied by the given system structure to pass the test. The evaluation of the constraints can be automatized by parsing them to checks whether or not a test succeeds. Besides this evaluation based on the constraints, {\framework} also supports smoke tests, e.g., observations whether the system throws exceptions during execution or simply crashes. This enables a completely automatized test process within the framework and addresses C-Oracle.
 
%For evaluation, the test oracle assesses the validity of the computed solution of the SOuT and, separately, the system structure that is present before the next test case is executed, i.e., the deployed systems structure. The latter case allows to evaluate the constraints of the system after a reorganization result is adopted by the agents.
%The computed solution within the SOuT, which are distributed afterwards, are likewise checked for violations by the test monitor.
%This gray-box-based evaluation can reveal failures within the computation of new system structures as well as within the application of reorganization results, consequently challenge~(C2) is addressed here.
As shown in \cref{fig:overviewFramework}, the monitoring and evaluation component generates test run results for a specific test suite. The test run result contains evaluations of each test case result (failed, passed) as well as the logged information (system configuration, SOuT solution) during the test run for each executed test case.

\section{Tested Self-organization Algorithms}\label{sec:testedSOalgorithms}

In this section, we present two SO algorithms, a decentralized approach (\cref{sec:spada}) and a metaheuristic (\cref{sec:pso}).
The aim of both algorithms is to partition a set of agents $\mathcal{A} = \{a_1, \dots, a_n\}$ into pairwise disjoint subsets, i.e., partitions, that together constitute a \emph{partitioning} at as minimal costs as possible.
With regard to our case study, each AVPP represents a partition and the set of all AVPPs corresponds to a partitioning.
As SPADA and \psosp\ are part of the controller of the CEI, the oracle has to evaluate if their interim results (R-Solution) as well as the resulting system structures (R-Distribution) meet the functional requirements, i.e., satisfy the properties of partitionings (cf.\ \cref{sec:evaluation}).
%\todo[inline]{Sounds to me like we don't address the other challenges, could we address them here?}
%Because both algorithms make use of randomized decisions to find high-quality solutions in large search spaces, a testing approach has to deal with very large state spaces (C-BranchingStateSpace).

\subsection{A Decentralized Algorithm for Partitioning Multi-Agent Systems}\label{sec:spada}
SPADA~\cite{anders2012spada}, the \emph{\textbf{S}et \textbf{P}artitioning \textbf{A}lgorithm for \textbf{D}istributed \textbf{A}gents}, solves the \emph{complete set partitioning problem}~(CSPP) in a general, decentralized manner.
%%In the CSPP, the goal is to partition a set $\mathcal{A} = \{a_1, \dots, a_n\}$ into pairwise disjoint subsets, i.e., partitions, that exhibit application-specific properties.
%Because SPADA allows the definition of application-specific metrics, it can be applied to a variety of problems.
%For instance, if the objective is to group similar or dissimilar elements together, the CSPP is equivalent to clustering or anticlustering~\cite{valev1998set}.
%In case a metric defines how well agents can work together on a common task, the CSPP is equivalent to coalition structure generation~\cite{RahwanRJG09}.
%%Since SPADA has been designed to solve the CSPP in general, it can be applied to these specific problems as well.
%This distinguishes SPADA from other centralized and decentralized approaches, which are often specialized to a specific problem in a specific domain.
In the following, we give a short summary of SPADA's basic functionality and characteristics.
A more detailed description can be found in \cite{anders2012spada}.
%We use the term ``reorganisation'' to denote the process performed by SPADA.

\begin{figure}[h]
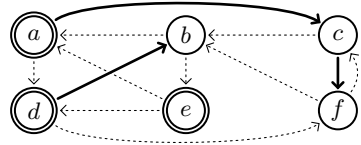

\centering
%%%%%%%%%%%%%%%Bild%%%%%%%%%%%%%%%
\begin{pgfpicture}{0cm}{0cm}{5.8cm}{1.75cm}
\begin{pgftranslate}{\pgfpoint{2.9cm}{1.5cm}} %Ursprung des Koordinatensystems oben Mitte
%%%Knoten
%\pgfrect[stroke]{\pgfpoint{-4.25cm}{0cm}}{\pgfpoint{8.5cm}{10pt}}
\pgfsetlinewidth{0.75pt}
\pgfnodecircle{Node1_1_r}[stroke]{\pgfxy(-2,0)}{0.25cm}
\pgfnodecircle{Node1_1}[stroke]{\pgfxy(-2,0)}{0.3cm}
\pgfnodecircle{Node1_2}[stroke]{\pgfrelative{\pgfxy(2,0)}{\pgfnodecenter{Node1_1}}}{0.25cm}
\pgfnodecircle{Node1_3}[stroke]{\pgfrelative{\pgfxy(4,0)}{\pgfnodecenter{Node1_1}}}{0.25cm}
\pgfnodecircle{Node2_1_r}[stroke]{\pgfrelative{\pgfxy(0,-1)}{\pgfnodecenter{Node1_1}}}{0.25cm}
\pgfnodecircle{Node2_1}[stroke]{\pgfrelative{\pgfxy(0,-1)}{\pgfnodecenter{Node1_1}}}{0.3cm}
\pgfnodecircle{Node2_2_r}[stroke]{\pgfrelative{\pgfxy(2,0)}{\pgfnodecenter{Node2_1}}}{0.25cm}
\pgfnodecircle{Node2_2}[stroke]{\pgfrelative{\pgfxy(2,0)}{\pgfnodecenter{Node2_1}}}{0.3cm}
\pgfnodecircle{Node2_3}[stroke]{\pgfrelative{\pgfxy(4,0)}{\pgfnodecenter{Node2_1}}}{0.25cm}
%%%Kanten
\pgfsetlinewidth{0.4pt}
\pgfnodesetsepstart{1.5pt} %Abstand vom Knoten
\pgfnodesetsepend{1.5pt}
\pgfsetendarrow{\pgfarrowlargepointed{2pt}} %mit Pfeilspitze
\pgfsetroundcap
%ungestrichelte, dicke Kanten
\pgfsetlinewidth{1.0pt} %Kante dick
\pgfnodeconncurve{Node1_1}{Node1_3}{35}{145}{0.6cm}{0.6cm}
\pgfnodeconnline{Node1_3}{Node2_3}
\pgfnodeconnline{Node2_1}{Node1_2}
%gestrichelte Kanten
\pgfsetlinewidth{0.4pt}
\pgfsetdash{{0.025cm}{0.05cm}}{0cm}
\pgfnodeconnline{Node1_1}{Node2_1}
\pgfnodeconnline{Node1_2}{Node2_2}
\pgfnodeconnline{Node1_2}{Node1_1}
\pgfnodeconnline{Node1_3}{Node1_2}
\pgfnodeconncurve{Node2_1}{Node2_3}{325}{215}{0.6cm}{0.6cm}
\pgfnodeconnline{Node2_2}{Node2_1}
\pgfnodeconnline{Node2_2}{Node1_1}
\pgfnodeconnline{Node2_3}{Node1_2}
\pgfnodeconncurve{Node2_3}{Node1_3}{55}{305}{0.25cm}{0.25cm}
%%%Beschriftungen
\pgfputat{\pgfrelative{\pgfxy(0,0)}{\pgfnodecenter{Node1_1}}}{\pgfbox[center,center]{$a$}}
\pgfputat{\pgfrelative{\pgfxy(0,0)}{\pgfnodecenter{Node1_2}}}{\pgfbox[center,center]{$b$}}
\pgfputat{\pgfrelative{\pgfxy(0,0)}{\pgfnodecenter{Node1_3}}}{\pgfbox[center,center]{$c$}}
\pgfputat{\pgfrelative{\pgfxy(0,0)}{\pgfnodecenter{Node2_1}}}{\pgfbox[center,center]{$d$}}
\pgfputat{\pgfrelative{\pgfxy(0,0)}{\pgfnodecenter{Node2_2}}}{\pgfbox[center,center]{$e$}}
\pgfputat{\pgfrelative{\pgfxy(0,0)}{\pgfnodecenter{Node2_3}}}{\pgfbox[center,center]{$f$}}
\end{pgftranslate}
\end{pgfpicture}
%\centering
%\caption{A system consisting of three partitions~$\{a,c,f\},\{b,d\},\{e\}$ with leaders~$a,d,e$. Dashed edges represent unmarked links, whereas solid edges represent marked links.}
\caption{An exemplary acquaintances graph for a system consisting of six agents (cf.~\cite{anders2012spada}): Agents are represented as nodes and acquaintances as directed edges, e.g., $d$ is acquainted with $b$ and $f$.
Marked edges (symbolized as solid arcs) indicate that their tail and head belong to the same partition.
In this example, there are three partitions~$\{a,c,f\},\{b,d\},\{e\}$ with leaders~$a,d,e$.}
\label{fig:spada-model}
\end{figure}

In SPADA, the agents use an internal graph-based representation of the current partitioning, called \emph{acquaintances graph}, to solve the CSPP.
All operations the agents apply to establish a suitable partitioning can therefore be mapped to graph operations.
The nodes of the acquaintances graph are the agents participating in the reorganization. Directed edges represent acquaintance relationships between agents.
Together the acquaintances form an overlay network that restricts communication to acquainted agents, thereby lowering complexity in large systems.
%To simplify the application of the graph operations that modify the partitioning, it is defined that each partition is respresented by a directed tree of marked links.
To indicate that an agent is not only acquainted with another but also in the same partition, edges can be marked.
Partitions are thus defined by the transitive-reflexive closure of the binary relation given by the marked edges.
Each partition has a designated leader that is responsible for optimizing its composition according to application-specific criteria.
%Hence, the whole partitioning is a directed forest.
An example of such an acquaintances graph is depicted in \cref{fig:spada-model} (more details concerning the acquaintances graph can be found in \cite{anders2012spada}).

%\begin{figure}[t]
	%\centering
		%\includegraphics[width=0.75\textwidth]{spada-graphs.pdf}
	%\caption{An instance of SPADA's acquaintance graph and three color-coded partitions. Unmarked edges are denoted by dashed arcs, marked edges are denoted by solid arcs. The leaders of the three partitions are identified with an ``x''.
	%%Please note that an actual graph would be more strongly connected than the one depicted here.
	%}
	%\label{fig:spada-graphs}
%\end{figure}

%Each partition has a designated leader, which is the root of the tree representing the corresponding partition. 
%Leaders are responsible for optimizing the composition of their partitions according to application-specific criteria.
%Among others, this includes managing the partition's composition.
%For doing so, it knows all partition members as well as their acquaintances.
To improve its partition, each leader periodically evaluates if it is beneficial to integrate new agents %into its partition
or to exclude some of its members (e.g., with regard to our case-study, to improve the equal distribution of unreliable power plants among AVPPs).
The latter can be beneficial in case of reorganizations that require to create new partitions, e.g., if a partition's or an agent's properties have changed so that the partition's formation criteria no longer favor including the agent.
The integration and exclusion of agents is implemented by modifying the edges in the acquaintances graph. %creating new and redirecting or dissolving existing edges in the acquaintances graph.
%Potential candidates for integration are acquainted agents that are not yet in the partition.

To decide about termination, leaders periodically evaluate ap\-pli\-ca\-tion-spe\-cif\-ic termination criteria.
These are formulated as constraints which can also be monitored at run-time to trigger reorganization.
%Therefore, their evaluation only necessitates the leader's local knowledge.
%These are formulated as predicates on the basis of the leaders' local knowledge.
%These predicates are constraints which, usually, are also monitored at runtime to trigger reorganization.
If the termination criteria are met, the leader marks its partition as terminated.
As long as a partition is marked as terminated, its leader does not change its structure.
However, the termination labeling is removed if the partition is changed from outside, i.e., if one of its members is integrated into another partition.
This characteristic allows SPADA to make selective changes to an existing partitioning, which is very useful in dynamic environments. 
It has been shown empirically that SPADA's local decisions lead to a partitioning whose quality is within $10\%$ of the optimum~\cite{anders2012spada}.

With regard to our case study, each leader instantiates a new AVPP agent as soon as all partitions terminated.
The AVPP then assumes control of all power plants in the partition.
In case of a reorganization, the acquaintances graph is created on the basis of the existing system structure.

%This means that each AVPP that takes part in the reorganization is represented by a tree of marked edges.
%The trees' structure as well as the unmarked edges are created at random such that each agent has the same number of outgoing edges, i.e., acquaintances.

\subsection{A Particle Swarm Optimizer for Partitioning Multi-Agent Systems}\label{sec:pso}
\emph{\psosp}~\cite{anders2015pso}, the \emph{\textbf{P}article \textbf{S}warm \textbf{O}pti\-mizer for the \textbf{P}ar\-ti\-tion\-ing \textbf{P}roblem}, is based on \emph{Particle Swarm Optimization}~(PSO)~\cite{kennedy95}, a bio-inspired computational method and metaheuristic for optimization in large search spaces. %for finding nearly optimal solutions in large search spaces in an efficient way.
In PSO, a number of particles concurrently explore the search space in search of better candidate solutions by modifying their current positions (at random or by approaching other candidate solutions) as long as a specific termination criterion is not met.
During this process, each particle's current position represents a specific candidate solution.
To be able to improve the quality of candidate solutions in a target-oriented manner, each particle~$\aParticle$ is aware of its best found solution $\bestParticle{i}$ and the best found solution $\bestNeighborhood{i}$ in its neighborhood $\neighborhood{i}$.
The algorithm's outcome is the global best found solution $\globalBest$.

\psosp\ solves a variant of the CSPP in the presence of \emph{partitioning constraints} that constrain feasible partitions in terms of a minimum $\minPartitionSize$ and a maximum $\maxPartitionSize$ size as well as a minimum $\minNbPartitions$ and a maximum $\maxNbPartitions$ number of partitions.
Therefore, the test oracle additionally has to check---without interfering the algorithm, by evaluating the logged data of the algorithm and not locking it during execution---if the interim results and the resulting system structure satisfy the partitioning constraints.
In \psosp, each particle represents a partitioning that satisfies the partitioning constraints.
%~$\aParticle$ $\partitioningA$ that consists of $\sizePartitioning{\partitioningA}$ partitions ($\minNbPartitions \leq \sizePartitioning{\partitioningA} \leq \maxNbPartitions$).
%In this partitioning, every partition $\partitionA \in \partitioningA$ consists of $\minPartitionSize \leq \sizePartition{\partitionA} \leq \maxPartitionSize$ elements.
%To differentiate this specific problem from the CSPP more clearly, we will refer to it as the partitioning problem~(PP).
%To differentiate this specific problem from the original SPP, in which feasible, i.e., valid, partitions are predefined, more clearly, we will refer to it as the partitioning problem~(PP).
%In the unbounded case, the PP corresponds to the CSPP.
%The application of a metaheuristic is suitable because of the PP's complexity.
%
%\psosp\ \enumb{1}~solves the PP in a general manner and \enumb{2}~allows to specify and efficiently deal with suitable ranges for the number as well as the size of partitions.
%\psosp\ solves the partitioning problem regarded in this paper in a general manner.
The central idea---which could also be applied to other metaheuristics---is to allow the particles to move around the search space by using the basic set operations \emph{join}, \emph{split}, and \emph{exchange}. %to come to a solution.
The join operation creates the union of two partitions, the split operation divides an existing partition into two non-empty subsets, and the exchange operation exchanges elements between two partitions. %and basically corresponds to a join that is followed by a split.
Particles can apply these operations at random as well as in a target-oriented manner.
The main purpose of the former case is to enable the particles to explore the search space by randomly modifying their represented partitioning, i.e., their position. %is modified by determining the partitions to change and the involved subsets at random.
The latter case, in contrast, allows particles to exploit existing candidate solutions by approaching other candidate solutions in promising regions of the search space.
%In the latter case, it is guaranteed that the similarity of the resulting partitioning to the approached partitioning is not lower than the similarity before applying the operator.
Since \psosp's operations are defined in a way that their application always maintains solution correctness, it combs through a search space that only contains correct solutions.
This is advantageous with regard to its performance. Because \psosp\ is initialized with a correct candidate solution, it is an any time~algorithm.

Similar to SPADA, \psosp\ can be customized to a specific application by devising an appropriate fitness function that assesses the quality of solutions and thus steers the search for them.
Due to these characteristics, \psosp\ can be applied to many different applications in which solving the partitioning problem considered in this paper is relevant and global knowledge is available.

Having specified valid partitionings by means of $\minNbPartitions, \maxNbPartitions, \minPartitionSize, \maxPartitionSize$ as well as the particles' attitude towards exploration and exploitation by fixing some parameters that influence the probability that particles make a random move or approach other candidate solutions, %by fixing the constants $c_\mathit{rdm}, c_{\bestParticle{i}}, c_{\bestNeighborhood{i}}$,
\psosp\ creates a predefined number of particles at random or predetermined positions in the search space (the set of particles does not change at run-time).
The latter is suitable when a reorganization of an existing system structure has to take place:
If the current structure does not contradict the partitioning constraints, it can be used as a starting point for the self-organization process. Mixing predefined and randomly generated initial partitionings allows to hold up diversity.
When searching for an initial system structure, %, i.e., partitioning,
particles are created at \mbox{random positions.}
%Since the boundaries for the number and size of partitions represent hard constraints, partitionings that do not meet them %are regarded as invalid and
%do not have to be represented in the search space and the time needed to find high-quality solutions can be lowered (cf.\ \todo{Evaluation}).
%Because of the reduction of the search space, suitable boundaries can lower the time needed to find high-quality solutions (cf.\ \todo{Evaluation}).

\begin{figure}[t]
		\centering
		 \includegraphics[width=0.85\textwidth]{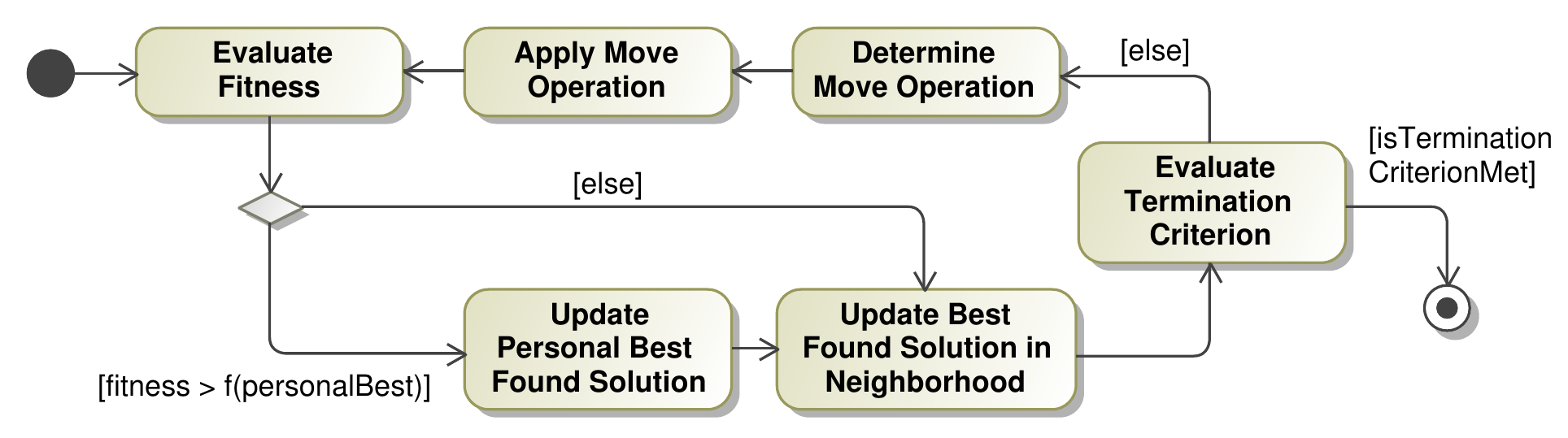}
			\caption{Actions performed by particles in each iteration~\cite{anders2015pso}.}
		\label{fig:procedure}
		\vspace{-0.4cm}
\end{figure}

%The position of each particle~$\aParticle$ represents a partitioning $\partitioningA$ that consists of $\sizePartitioning{\partitioningA}$ partitions ($\minNbPartitions \leq \sizePartitioning{\partitioningA} \leq \maxNbPartitions$).
%In this partitioning, every partition $\partitionA \in \partitioningA$ consists of $\minPartitionSize \leq \sizePartition{\partitionA} \leq \maxPartitionSize$ elements.
%All particles concurrently explore the search space in search of better solutions by modifying their current positions (at random or by approaching other solutions) as long as a specific termination criterion is not met.
As long as a specific termination criterion is not met, a particle $\aParticle$ performs the following actions in each iteration; these are also depicted in \cref{fig:procedure}:
\begin{enumerate}[leftmargin=*]
	\item Evaluate the fitness $\fitness{\partitioningA}$ of the represented partitioning~$\partitioningA$. %The fitness function corresponds to the ``metric'' used in the PP's definition in \cref{sec:introduction}.
	\item If the particle's fitness $\fitness{\partitioningA}$ is higher than the fitness $\fitness{\bestParticle{i}}$ of its best found solution $\bestParticle{i}$, set $\bestParticle{i}$ to~$\partitioningA$. Further, inform all other particles $\bParticle$ that contain $\aParticle$ in their neighborhood $\neighborhood{j}$ about the improvement so that they can update $\bestNeighborhood{j}$, i.e., the best found solution in their neighborhood.
	\item Update the best found solution~$\bestNeighborhood{i}$ in $\aParticle$'s neighborhood~$\neighborhood{i}$.
	\item Stop if the termination criterion is met.
	\item Otherwise, randomly opt for the direction in which to move, i.e., choose whether a random move or an approach operation should be applied. In case of an approach operation, also determine the position (i.e., $\bestParticle{i}$ or $\bestNeighborhood{i}$) that should be approached.
	\item Determine the new position $\resultingPartitioning{\partitioningA}$ by applying the selected move operation to~$\partitioningA$.
\end{enumerate}
Once all particles terminated, \psosp\ returns the best found solution~$\globalBest$.
Possible termination criteria are, e.g., a predefined amount of time, a predefined number of iterations (i.e., moves through the search space), a predefined threshold for the minimum fitness value, or a combination of these criteria.

\section{Evaluation}\label{sec:evaluation}

In our evaluation, we used a Java implementation of \framework\ that is based on a multi-agent system called TEMAS~\cite{temas2013}.
We opted for TEMAS because it supports a stepwise execution out of the box, which allows \framework\ to monitor consistent states of the system at specific points in time.

As foundation of the \emph{model of the system under test}, we used the domain model of the AVPP application, which can be found in~\cite{refarch2013}.
The model of the environmental changes and influences described the effect of different environmental conditions, such as weather conditions, on the power plants' ability to make adequate predictions of their future output.
Clearly, this effect depends on the concrete type of agent, i.e., power plant. For this reason, we regarded four different agent types: solar panels, wind turbines, biogas power plants, and hydro power plants. Based on the assumption that the effect of changing environmental conditions, such as the global radiation, the wind speed, the available amount of biogas, or the water flow, is characteristic of a specific type of power plant, we generated different sets of agent groups.
Further, the power plants' geographic location was taken into account.
Considering an AVPP's prediction accuracy as a property resulting from the average prediction accuracy of its members, the system's goal was to maintain a structure of AVPPs that feature a similar prediction accuracy (cf.\ \cref{sec:AVPP}).
As soon as the dissimilarity of the AVPPs' prediction accuracy exceeded a certain threshold (as a result of environmental changes), the power plants triggered a reorganization that should reestablish the similarity.
We parametrized the \emph{model of the system under test} as follows:
\begin{itemize}[leftmargin=*]
	\item number of agents $\#\mathit{ag}$: between 2 and 1000
	\item agent group size: between 2 and $\#\mathit{ag}$
	\item number of agent groups: between 1 and $\big\lfloor \frac{1}{2}\cdot\#\mathit{ag} \big\rfloor$
	\item partition size: $2 \leq \minPartitionSize \leq \maxPartitionSize \leq \#\mathit{ag}$
	\item number of partitions: $1 \leq \minNbPartitions \leq \maxNbPartitions \leq \big\lfloor \frac{1}{2}\cdot\#\mathit{ag} \big\rfloor$
	\item 10 test sequences per test suite
	\item number of test cases per test sequence: between 50 and 1000
	\item number of states per EP: between 3 and 25
\end{itemize}

As SOuT, we integrated a Java-based implementation of the partitioning algorithms SPADA and \psosp\ (cf.\ \cref{sec:testedSOalgorithms}).
Both implement the IController interface that is used by the SO Algorithm Adapter to initiate the SOuT, request results, and ask the SOuT to adopt the new system structure (cf.\ \cref{fig:Component_Diagram__Testsystem}).
The latter was implemented by sequentially moving the power plants contained in the calculated partitioning from their current AVPP into the corresponding new AVPP.
With regard to the system structure, this procedure assures that every power plant is always contained in exactly one AVPP.
If the last power plant was removed from an AVPP, this AVPP was dissolved.
The description of the algorithms provided in \cite{anders2012spada} and \cite{anders2015pso} as well as their implementation served as the \emph{model of the SOuT}. As explained in \cref{sec:testModel}, we used this information to identify relevant parameters and suitable valid ranges for their parametrization.
For SPADA and \psosp, we identified the following parameters:
\begin{itemize}[leftmargin=*]
	\item SPADA
	\begin{itemize}[label={$\circ$}, leftmargin=*]
		\item number of acquaintances per agent: between $1$ and $20$
		\item number of agents each leader evaluates for integration into its partition: between $1$ and $10$
		\item maximum number of agents a leader can integrate into its partition within a single step: between $1$ and $10$
	\end{itemize}
	\item \psosp
	\begin{itemize}[label={$\circ$}, leftmargin=*]
		\item number of particles $\#\mathit{P}$: between 1 and 4
		\item number of particles starting at the current partitioning: between 0 and~$\#\mathit{P}$
		\item probabilities $c_{\mathit{rdm}}, c_{\bestParticle{i}}, c_{\globalBest} \in [0,1]$ (with $c_{\mathit{rdm}} + c_{\bestParticle{i}} + c_{\globalBest} = 1$) to apply a random move operator, approach the particle's best found solution $\bestParticle{i}$, and approach the global best found solution~$\globalBest$, respectively
		\item max. run-time in seconds: between $1$ and $10$
	\end{itemize}
\end{itemize}

After each reorganization, the oracle checks if the algorithm's result complies with the definition of partitionings (i.e., each power plant must be a member of exactly one AVPP).
As explained in \cref{subsec:executionComponent} and \cref{sec:outputComponent}, these checks are performed once the algorithm indicates its termination (R-Solution) as well as after the result has been adopted (R-Distribution).
Only in case of \psosp, the oracle additionally evaluated the satisfaction of the partitioning constraints introduced in \cref{sec:pso} since SPADA does not allow to restrict valid partitionings with regard to the size and number of partitions.

\subsection{Fault Injection}\label{sec:fault-injection}
To evaluate our approach, we injected four faults into the SPADA and five faults into the \psosp\ implementation.
According to P\"uschel et al.~\cite{puschel2014testing}, all these faults can be assigned to the classes of ``permanent and transient RECONF faults''.
Our injected faults have in common that they do not cause the algorithm to throw an exception that simply has to be caught by the oracle (i.e., smoke tests), but that their application can result in an invalid reorganization result or invalid system structure.
In a preliminary evaluation that ran for about one week, we tested the SPADA and \psosp\ implementation without injecting any faults.
We did not observe any failures in the course of these tests.
So we can be confident that the failures the oracle reported during our subsequent evaluation can be attributed to a specific injected fault.

\paragraph{SPADA: Injected Faults.}
The first two types of SPADA faults (cf.\ SPADA-F1 and SPADA-F2) manifest in an incorrect transformation of the current system structure into SPADA's internal model of a partitioning, i.e., the acquaintances graph (cf.\ \cref{sec:spada}). This false mapping results in an invalid reorganization result.

\begin{description}[font=\rmfamily\itshape\mdseries, leftmargin=0cm, style=nextline]
	\item[SPADA-F1/SPADA-F2]
\begin{itemize}[leftmargin=*, midpenalty=1000]\item[]
	\item \emph{Description}: When creating the acquaintances graph for a new reorganization on the basis of the current system structure, an arbitrary AVPP is not represented in the acquaintances graph if the number of AVPPs is above (in case of SPADA-F1) or below (in case of SPADA-F2) a certain threshold. We set these thresholds to 100 for SPADA-F1 and to 5 for SPADA-F2.
	\item \emph{Effect}: The resulting partitioning does not contain the power plants that have been members of the ``forgotten'' AVPP.
	%\item[Difficulty of Discovery:] easy because ... (parameters arbitrary, sufficient number of repetitions)
\end{itemize}
\end{description}

The two other types of faults we integrated into SPADA concern a functionality that is used to transform the result, given in the form of an acquaintances graph, into a set of sets.
This functionality is used to provide the result to the SO Algorithm Adapter (R-Solution) and as a preprocessing step to create the new AVPP structure (R-Distribution).

\begin{description}[font=\rmfamily\itshape\mdseries,
                    leftmargin=0cm,
                    itemsep=8pt,
                    style=nextline]
	\item[SPADA-F3]
\begin{itemize}[leftmargin=*, midpenalty=1000, itemsep=0pt]\item[]
	\item \emph{Description}: In case the size of a partition exceeds a predefined threshold, arbitrary power plants are deleted from this partition until its size equals this threshold. In our evaluation, we set this threshold to 100.
	\item \emph{Effect}: Some power plants are not represented in the partitioning.
	%\item[Difficulty of Discovery:] hard because ... (concrete parametrization needed)
\end{itemize}

	\item[SPADA-F4]
\begin{itemize}[leftmargin=*, midpenalty=1000, itemsep=0pt]\item[]
	\item \emph{Description}: In case the size of a partition exceeds a predefined threshold, this partition is replaced by a partition that is randomly selected from the partitioning. In our evaluation, we set this threshold to 100.
	\item \emph{Effect}: Some power plants are not represented in the partitioning, whereas others occur two or more times.
	%\item[Difficulty of Discovery:] hard because ... (concrete parametrization needed)
\end{itemize}
\end{description}

All SPADA faults can be detected using the gray-box interface (R-Solution), i.e., before the underlying system structure is changed.
Given the way the result is transformed into a new system structure (it is ensured that every power plant is always a member of exactly one AVPP), these faults \emph{cannot} be detected using the black-box view (R-Distribution). Note that the oracle does not check the system structure with respect to the number and the size of AVPPs in case of SPADA.
For each type of injected fault, we are thus confronted with the problem of error masking (C-ErrorMask).

\paragraph{\psosp: Injected Faults.}
Regarding \psosp, we modified the implementation of the move operations ``random split'' (\psosp-F1), ``random join'' (\psosp-F2), ``approach split'' (\psosp-F3), ``approach join'' (\psosp-F4), and ``approach exchange'' (\psosp-F5) as described in the following listing.

\begin{description}[font=\rmfamily\itshape\mdseries,
                    leftmargin=0cm,
                    itemsep=8pt,
                    style=nextline]
	\item[\psosp-F1]
\begin{itemize}[leftmargin=*, midpenalty=1000, itemsep=0pt]\item[]
	\item \emph{Description}: If a partition $\partitionA$ is randomly split into two partitions $\partitionB$ and $\partitionC$, an arbitrary power plant of $\partitionB$ is replaced by another arbitrary power plant of $\partitionC$. This fault does only occur if the size of $\partitionB$ and $\partitionC$ is below a threshold~$t_1$ or above a threshold~$t_2$.
	\item \emph{Effect}: With regard to the resulting partitioning, a specific power plant is missing and another occurs twice.
	%\item[Difficulty of Discovery:] hard because ...
\end{itemize}

	\item[\psosp-F2]
\begin{itemize}[leftmargin=*, midpenalty=1000, itemsep=0pt]\item[]
	\item \emph{Description:} If two partitions $\partitionA$ and $\partitionB$ are merged into a new partition $\partitionC$ when applying the random join operator, either $\partitionA$ or $\partitionB$ is not removed from the partitioning.
	This fault does only occur if the size of $\partitionA$ and $\partitionB$ is below $t_1$ or above $t_2$.
	\item \emph{Effect:} In the resulting partitioning, the power plants of either $\partitionA$ or $\partitionB$ occur twice as they are also contained in $\partitionC$.
	%\item[Difficulty of Discovery:] medium because ... (specific range of parameters needed)
\end{itemize}

	\item[\psosp-F3]
\begin{itemize}[leftmargin=*, midpenalty=1000, itemsep=0pt]\item[]
	\item \emph{Description:} If a partition $\partitionA$ is split into two partitions $\partitionB$ and $\partitionC$, the resulting partitioning does not contain either $\partitionB$ or $\partitionC$. This fault does only occur if the size of $\partitionB$ and $\partitionC$ is below $t_1$ or above $t_2$.
	\item \emph{Effect:} In the resulting partitioning, the power plants of either partition $\partitionB$ or $\partitionC$ \mbox{are missing.}
	%\item[Difficulty of Discovery:] medium because ... (specific range of parameters needed)
\end{itemize}

	\item[\psosp-F4]
\begin{itemize}[leftmargin=*, midpenalty=1000, itemsep=0pt]\item[]
	\item \emph{Description}: If two partitions $\partitionA$ and $\partitionB$ are merged into a new partition $\partitionC$ when applying the approach join operator, one element is removed from $\partitionC$.
	This fault does only occur if the size of $\partitionA$ and $\partitionB$ is below $t_1$ or above $t_2$.
	\item \emph{Effect}: In the resulting partitioning, a single power plant is missing.
	%\item[Difficulty of Discovery:] medium because ... (specific range of parameters needed)
\end{itemize}

	\item[\psosp-F5]
\begin{itemize}[leftmargin=*, midpenalty=1000, itemsep=0pt]\item[]
	\item \emph{Description}: If some power plants are exchanged between two partitions $\partitionA$ and $\partitionB$, one power plant of either $\partitionA$ or $\partitionB$ occurs in both resulting partitions $\partitionC$ and $\partitionD$. This fault does only occur if the size of $\partitionC$ and $\partitionD$ is below $t_1$ or above $t_2$.
	\item \emph{Effect}: In the resulting partitioning, one power plants occurs twice.
	%\item[Difficulty of Discovery:] medium because ... (specific range of parameters needed)
\end{itemize}
\end{description}

In our experiments, we used $t_1 = 2$ and $t_2 = 100$ so that the failures do only occur in certain situations.
Note that the application of an injected fault does not necessarily yield an invalid result because an invalid candidate solution must be rated better than all other (possibly valid) candidate solutions found by the particles (C-ErrorMask).
Clearly, this characteristic together with \psosp's non-deterministic behavior exacerbates the detection of an injected fault (C-BranchingStateSpace).

In principle, all types of \psosp\ faults could lead to a false result that can be detected using the gray-box interface (R-Solution) as well as the black-box view (R-Distribution).
Note, however, that not all invalid results manifest themselves in an invalid system structure.
Consider the following example illustrating error masking (C-ErrorMask): Assume that the power plants $a$ and $b$ are currently members of the same AVPP. If a reorganization causes an invalid result that does not contain these two power plants, the oracle detects a failure using the gray-box interface. However, the resulting system structure is \emph{valid} in case the minimum size of an AVPP is $\leq 2$ and the maximum number of AVPPs is not exceeded.
This is because $a$ and $b$ simply remain in their old AVPP if they are not contained in the provided result.

\subsection{Test Execution}

\begin{table}[!t]
	\centering
	{
		\def\tempwidth{0.85cm} %width of the multirow (-> for line breaking) of all values in the table that use multirows
		\scriptsize
		\begin{tabularx}{\textwidth}{p{2.25cm}||>{\centering\arraybackslash}X|>{\centering\arraybackslash}X|>{\centering\arraybackslash}X|>{\centering\arraybackslash}X|>{\centering\arraybackslash}X|>{\centering\arraybackslash}X||>{\centering\arraybackslash}X|>{\centering\arraybackslash}X|>{\centering\arraybackslash}X|>{\centering\arraybackslash}X}
			\multirow{3}{*}{\parbox{2.0cm}{\raggedright Injected Fault}} & \multicolumn{6}{c||}{\psosp} & \multicolumn{4}{c}{SPADA} \\
			& F1 & F2 & F3 & F4 & F5 & F5d & F1 & F2 & F3 & F4 \\
			\hline\hline
			\raggedright \multirow{2}{*}{\#EP States} \phantom{xxxxxxxx}				& \multirow{2}{\tempwidth}{\raggedleft 11.68 \tiny{(5.39)}} & \multirow{2}{\tempwidth}{\raggedleft 15.77 \tiny{(6.49)}} & \multirow{2}{\tempwidth}{\raggedleft 12.30 \tiny{(5.96)}} & \multirow{2}{\tempwidth}{\raggedleft 13.74 \tiny{(5.75)}} & \multirow{2}{\tempwidth}{\raggedleft 15.41 \tiny{(6.16)}} & \multirow{2}{\tempwidth}{\raggedleft 14.25 \tiny{(6.05)}} & \multirow{2}{\tempwidth}{\raggedleft 16.42 \tiny{(5.71)}} & \multirow{2}{\tempwidth}{\raggedleft 14.72 \tiny{(6.42)}} & \multirow{2}{\tempwidth}{\raggedleft 16.62 \tiny{(6.19)}} & \multirow{2}{\tempwidth}{\raggedleft 15.44 \tiny{(6.31)}} \\
			\hline
			\raggedright \multirow{2}{*}{\#EP Transitions}	\phantom{xxxxxxxx}	& \multirow{2}{\tempwidth}{\raggedleft 127.42 \tiny{(99.20)}} & \multirow{2}{\tempwidth}{\raggedleft 222.06 \tiny{(148.59)}} & \multirow{2}{\tempwidth}{\raggedleft 143.31 \tiny{(124.31)}} & \multirow{2}{\tempwidth}{\raggedleft 169.78 \tiny{(128.68)}} & \multirow{2}{\tempwidth}{\raggedleft 210.22 \tiny{(142.27)}} & \multirow{2}{\tempwidth}{\raggedleft 183.28 \tiny{(130.00)}} & \multirow{2}{\tempwidth}{\raggedleft 230.83 \tiny{(134.75)}} & \multirow{2}{\tempwidth}{\raggedleft 197.09 \tiny{(134.29)}} & \multirow{2}{\tempwidth}{\raggedleft 240.56 \tiny{(141.24)}} & \multirow{2}{\tempwidth}{\raggedleft 212.72 \tiny{(136.51)}} \\
			\hline
			\raggedright \%EP State Coverage					& \multirow{2}{\tempwidth}{\raggedleft 99.51 \tiny{(2.33)}} & \multirow{2}{\tempwidth}{\raggedleft 99.57 \tiny{(2.08)}} & \multirow{2}{\tempwidth}{\raggedleft 99.45 \tiny{(2.50)}} & \multirow{2}{\tempwidth}{\raggedleft 99.42 \tiny{(2.71)}} & \multirow{2}{\tempwidth}{\raggedleft 99.47 \tiny{(2.35)}} & \multirow{2}{\tempwidth}{\raggedleft 99.54 \tiny{(2.31)}} & \multirow{2}{\tempwidth}{\raggedleft 99.66 \tiny{(1.77)}} & \multirow{2}{\tempwidth}{\raggedleft 99.59 \tiny{(2.15)}} & \multirow{2}{\tempwidth}{\raggedleft 99.54 \tiny{(2.35)}} & \multirow{2}{\tempwidth}{\raggedleft 99.60 \tiny{(1.98)}} \\
			\hline
			\raggedright \%EP Transition Coverage			& \multirow{2}{\tempwidth}{\raggedleft 52.31 \tiny{(15.80)}} & \multirow{2}{\tempwidth}{\raggedleft 45.37 \tiny{(12.63)}} & \multirow{2}{\tempwidth}{\raggedleft 50.57 \tiny{(14.32)}} & \multirow{2}{\tempwidth}{\raggedleft 47.53 \tiny{(11.97)}} & \multirow{2}{\tempwidth}{\raggedleft 44.76 \tiny{(11.47)}} & \multirow{2}{\tempwidth}{\raggedleft 47.57 \tiny{(14.11)}} & \multirow{2}{\tempwidth}{\raggedleft 43.98 \tiny{(11.34)}} & \multirow{2}{\tempwidth}{\raggedleft 47.46 \tiny{(15.11)}} & \multirow{2}{\tempwidth}{\raggedleft 43.52 \tiny{(11.60)}} & \multirow{2}{\tempwidth}{\raggedleft 45.34 \tiny{(13.23)}} \\
		\end{tabularx}
		}
	\caption{Statistical data concerning the number of EP states, EP transitions, as well as the coverage of EP states and EP transitions. All values are averages over the 700 generated test sequences per injected fault type. Values in parentheses denote standard~deviations.}
	\label{tab:evaluation-results-ep}
\end{table}
To be able to make a clear statement which types of faults can be found by \framework, only one specific type was injected during the execution of a single test sequence.
All in all, we injected 10 different types of faults: SPADA-F1 to SPADA-F4, \psosp-F1 to \psosp-F5, and an additional variant of \psosp-F5, called \psosp-F5d, which we will explain in more detail in the course of \cref{sec:eval-results}.
For each type, we generated 70 test suites, each containing 10 test sequences, resulting in 700 test sequences per fault type and a total number of executed test sequences of 7000.
Overall, we generated 3,679,326 test cases, corresponding to an average of 367,932.60 per fault type.
As shown in \cref{tab:evaluation-results-ep}, this high number of test cases allowed us to obtain an EP state coverage of more than 99\% and an EP transition coverage ranging between approximately 44\% and 52\% on average for all fault types.
To illustrate the advantages of the gray-box view, we did not abort the execution of a test sequence until the oracle found a failure using the black-box view.
In the course of the execution of a single test sequence, \framework\ could therefore register multiple failures using the gray-box interface.
On the other hand, it is not possible to detect more than one failure per test sequence using the black-box view.

\subsection{Evaluation Results}
\label{sec:eval-results}

As \cref{tab:evaluation-results-faults} shows, {\framework} was able to detect every kind of injected fault and
our evaluation results support our claims made in \cref{sec:fault-injection}: (1)~All failures detected using the black-box view are also detected using the gray-box view, and (2)~the SPADA faults cannot be detected using the black-box view.
The fact that not all \psosp\ faults that were disclosed using the gray-box view were also registered using the black-box view (between 0.00\% and 34.44\% on average) also demonstrates the problem of error masking (C-ErrorMask) in SOAS.
Except for \psosp-F4, the percentage of detected failures using the gray-box view (between 50.00\% and 82.31\% on average) outmatches the percentage of detected failures using the black-box view in all cases (between 31.25\% and 80.00\% on average).
These observations clearly indicate the need for gray-box interfaces in the context of testing SO algorithms.

\begin{table}[t!]
	\centering
	{
		\def\tempwidth{0.85cm} %width of the multirow (-> for line breaking) of all values in the table that use multirows
		\scriptsize
		\begin{tabularx}{\textwidth}{p{2.25cm}||>{\centering\arraybackslash}X|>{\centering\arraybackslash}X|>{\centering\arraybackslash}X|>{\centering\arraybackslash}X|>{\centering\arraybackslash}X|>{\centering\arraybackslash}X||>{\centering\arraybackslash}X|>{\centering\arraybackslash}X|>{\centering\arraybackslash}X|>{\centering\arraybackslash}X}
			\multirow{3}{*}{\parbox{2.0cm}{\raggedright Injected Fault}} & \multicolumn{6}{c||}{\psosp} & \multicolumn{4}{c}{SPADA} \\
%			& \multicolumn{6}{c||}{} & \multicolumn{4}{c}{} \\
			& F1 & F2 & F3 & F4 & F5 & F5d & F1 & F2 & F3 & F4 \\
			\hline\hline
			\raggedright \multirow{2}{*}{\#Agents}	\phantom{xxxxxxxxx}
			& \multirow{2}{\tempwidth}{\raggedleft 523.53 \tiny{(310.53)}} & \multirow{2}{\tempwidth}{\raggedleft 529.66 \tiny{(292.23)}} & \multirow{2}{\tempwidth}{\raggedleft 473.29 \tiny{(269.35)}} & \multirow{2}{\tempwidth}{\raggedleft 511.83 \tiny{(287.62)}} & \multirow{2}{\tempwidth}{\raggedleft 491.26 \tiny{(300.28)}} & \multirow{2}{\tempwidth}{\raggedleft 477.64 \tiny{(294.48)}}
			& \multirow{2}{\tempwidth}{\raggedleft 520.29 \tiny{(286.29)}} & \multirow{2}{\tempwidth}{\raggedleft 509.63 \tiny{(289.01)}} & \multirow{2}{\tempwidth}{\raggedleft 494.90 \tiny{(297.69)}} & \multirow{2}{\tempwidth}{\raggedleft 447.40 \tiny{(279.09)}} \\
			\hline
			\raggedright \multirow{2}{*}{\#Agent groups}	\phantom{xxxxxxxxx}
			& \multirow{2}{\tempwidth}{\raggedleft 39.80 \tiny{(67.18)}} & \multirow{2}{\tempwidth}{\raggedleft 37.42 \tiny{(48.83)}} & \multirow{2}{\tempwidth}{\raggedleft 20.09 \tiny{(27.72)}} & \multirow{2}{\tempwidth}{\raggedleft 36.05 \tiny{(64.05)}} & \multirow{2}{\tempwidth}{\raggedleft 30.2 \tiny{(47.79)}} & \multirow{2}{\tempwidth}{\raggedleft 32.74 \tiny{(48.77)}}
			& \multirow{2}{\tempwidth}{\raggedleft 31.86 \tiny{(57.70)}} & \multirow{2}{\tempwidth}{\raggedleft 35.12 \tiny{(47.42)}} & \multirow{2}{\tempwidth}{\raggedleft 31.30 \tiny{(58.27)}} & \multirow{2}{\tempwidth}{\raggedleft 38.85 \tiny{(45.35)}} \\
			\hline
			\raggedright \%Test sequences without failure
			& \multirow{2}{\tempwidth}{\raggedleft 79.97 \tiny{(40.05)}} & \multirow{2}{\tempwidth}{\raggedleft 84.55 \tiny{(36.17)}} & \multirow{2}{\tempwidth}{\raggedleft 91.42 \tiny{(28.03)}} & \multirow{2}{\tempwidth}{\raggedleft 96.28 \tiny{(18.94)}} & \multirow{2}{\tempwidth}{\raggedleft 97.86 \tiny{(14.49)}} & \multirow{2}{\tempwidth}{\raggedleft 70.22 \tiny{(45.76)}}
			& \multirow{2}{\tempwidth}{\raggedleft 97.13 \tiny{(16.69)}} & \multirow{2}{\tempwidth}{\raggedleft 84.22 \tiny{(36.48)}} & \multirow{2}{\tempwidth}{\raggedleft 85.69 \tiny{(35.04)}} & \multirow{2}{\tempwidth}{\raggedleft 90.13 \tiny{(29.85)}} \\
			\hline
			\raggedright \#Test cases per test sequence
			& \multirow{2}{\tempwidth}{\raggedleft 520.01 \tiny{(281.50)}} & \multirow{2}{\tempwidth}{\raggedleft 523.71 \tiny{(275.73)}} & \multirow{2}{\tempwidth}{\raggedleft 531.53 \tiny{(273.61)}} & \multirow{2}{\tempwidth}{\raggedleft 521.19 \tiny{(278.14)}} & \multirow{2}{\tempwidth}{\raggedleft 517.23 \tiny{(272.25)}} & \multirow{2}{\tempwidth}{\raggedleft 518.54 \tiny{(270.26)}}
			& \multirow{2}{\tempwidth}{\raggedleft 548.38 \tiny{(278.18)}} & \multirow{2}{\tempwidth}{\raggedleft 535.62 \tiny{(271.43)}} & \multirow{2}{\tempwidth}{\raggedleft 511.46 \tiny{(277.29)}} & \multirow{2}{\tempwidth}{\raggedleft 518.51 \tiny{(282.27)}} \\	
			\hline	
			\raggedright \%Applied test cases per test sequence
			& \multirow{3}{\tempwidth}{\raggedleft 81.37 \tiny{(38.39)}} & \multirow{3}{\tempwidth}{\raggedleft 85.69 \tiny{(34.44)}} & \multirow{3}{\tempwidth}{\raggedleft 92.74 \tiny{(25.62)}} & \multirow{3}{\tempwidth}{\raggedleft 96.77 \tiny{(17.11)}} & \multirow{3}{\tempwidth}{\raggedleft 98.62 \tiny{(11.16)}} & \multirow{3}{\tempwidth}{\raggedleft 73.90 \tiny{(42.75)}}
			& \multirow{3}{\tempwidth}{\raggedleft 100.00 \tiny{(0.00)}} & \multirow{3}{\tempwidth}{\raggedleft 100.00 \tiny{(0.00)}} & \multirow{3}{\tempwidth}{\raggedleft 100.00 \tiny{(0.00)}} & \multirow{3}{\tempwidth}{\raggedleft 100.00 \tiny{(0.00)}} \\
			\hline
			\raggedright \#Reorganizations per test sequence
			& \multirow{2}{\tempwidth}{\raggedleft 73.29 \tiny{(64.40)}} & \multirow{2}{\tempwidth}{\raggedleft 77.68 \tiny{(62.08)}} & \multirow{2}{\tempwidth}{\raggedleft 91.87 \tiny{(59.90)}} & \multirow{2}{\tempwidth}{\raggedleft 91.87 \tiny{(60.24)}} & \multirow{2}{\tempwidth}{\raggedleft 90.01 \tiny{(56.36)}} & \multirow{2}{\tempwidth}{\raggedleft 53.35 \tiny{(61.19)}}
			& \multirow{2}{\tempwidth}{\raggedleft 100.00 \tiny{(58.20)}} & \multirow{2}{\tempwidth}{\raggedleft 87.64 \tiny{(62.46)}} & \multirow{2}{\tempwidth}{\raggedleft 83.60 \tiny{(60.35)}} & \multirow{2}{\tempwidth}{\raggedleft 87.75 \tiny{(61.30)}} \\
			\hline
			\raggedright \#Failures per test sequence
			& \multirow{2}{\tempwidth}{\raggedleft 0.36 \tiny{(1.89)}} & \multirow{2}{\tempwidth}{\raggedleft 0.19 \tiny{(0.53)}} & \multirow{2}{\tempwidth}{\raggedleft 0.13 \tiny{(0.52)}} & \multirow{2}{\tempwidth}{\raggedleft 2.29 \tiny{(16.03)}} & \multirow{2}{\tempwidth}{\raggedleft 0.02 \tiny{(0.15)}} & \multirow{2}{\tempwidth}{\raggedleft 0.54 \tiny{(1.03)}}
			& \multirow{2}{\tempwidth}{\raggedleft 0.03 \tiny{(0.17)}} & \multirow{2}{\tempwidth}{\raggedleft 0.16 \tiny{(0.36)}} & \multirow{2}{\tempwidth}{\raggedleft 83.49 \tiny{(60.39)}} & \multirow{2}{\tempwidth}{\raggedleft 0.12 \tiny{(0.35)}} \\
			\hline
			\raggedright \#Failures
			& 252 & 130 & 91 & 1603 & 16 & 376
			& 20 & 110 & 58443 & 84 \\
			\hline
			\raggedright \%Undetected failures
			& \multirow{2}{\tempwidth}{\raggedleft 44.44} & \multirow{2}{\tempwidth}{\raggedleft 17.69} & \multirow{2}{\tempwidth}{\raggedleft 34.07} & \multirow{2}{\tempwidth}{\raggedleft 97.75} & \multirow{2}{\tempwidth}{\raggedleft 50.00} & \multirow{2}{\tempwidth}{\raggedleft 40.96}
			& \multirow{2}{\tempwidth}{\raggedleft 0.00} & \multirow{2}{\tempwidth}{\raggedleft 0.00} & \multirow{2}{\tempwidth}{\raggedleft 99.84} & \multirow{2}{\tempwidth}{\raggedleft 9.52} \\
			\hline
			\raggedright \%Detected failures (gray box)
			& \multirow{3}{\tempwidth}{\raggedleft 55.56} & \multirow{3}{\tempwidth}{\raggedleft 82.31} & \multirow{3}{\tempwidth}{\raggedleft 65.93} & \multirow{3}{\tempwidth}{\raggedleft 2.25} & \multirow{3}{\tempwidth}{\raggedleft 50.00} & \multirow{3}{\tempwidth}{\raggedleft 59.04}
			& \multirow{3}{\tempwidth}{\raggedleft 100.00} & \multirow{3}{\tempwidth}{\raggedleft 100.00} & \multirow{3}{\tempwidth}{\raggedleft 0.16} & \multirow{3}{\tempwidth}{\raggedleft 90.48} \\
			\hline
			\raggedright \%Detected failures (black box)
			& \multirow{3}{\tempwidth}{\raggedleft 53.17} & \multirow{3}{\tempwidth}{\raggedleft 80.00} & \multirow{3}{\tempwidth}{\raggedleft 57.14} & \multirow{3}{\tempwidth}{\raggedleft 2.25} & \multirow{3}{\tempwidth}{\raggedleft 31.25} & \multirow{3}{\tempwidth}{\raggedleft 51.33}
			& \multirow{3}{\tempwidth}{\raggedleft 0.00} & \multirow{3}{\tempwidth}{\raggedleft 0.00} & \multirow{3}{\tempwidth}{\raggedleft 0.00} & \multirow{3}{\tempwidth}{\raggedleft 0.00} \\
			\hline
			\raggedright \%Test sequences with failures detected in gray box only
			& \multirow{4}{\tempwidth}{\raggedleft 4.29 \tiny{(20.33)}} & \multirow{4}{\tempwidth}{\raggedleft 3.70 \tiny{(18.97)}} & \multirow{4}{\tempwidth}{\raggedleft 13.33 \tiny{(34.28)}} & \multirow{4}{\tempwidth}{\raggedleft 0.00 \tiny{(0.00)}} & \multirow{4}{\tempwidth}{\raggedleft 34.44 \tiny{(45.63)}} & \multirow{4}{\tempwidth}{\raggedleft 9.74 \tiny{(27.63)}}
			& \multirow{4}{\tempwidth}{\raggedleft 100.00 \tiny{(0.00)}} & \multirow{4}{\tempwidth}{\raggedleft 100.00 \tiny{(0.00)}} & \multirow{4}{\tempwidth}{\raggedleft 100.00 \tiny{(0.00)}} & \multirow{4}{\tempwidth}{\raggedleft 100.00 \tiny{(0.00)}} \\
			\hline
			\raggedright \%Test sequences with failures detected in black box only
			& \multirow{4}{\tempwidth}{\raggedleft 0.00 \tiny{(0.00)}} & \multirow{4}{\tempwidth}{\raggedleft 0.93 \tiny{(9.62)}} & \multirow{4}{\tempwidth}{\raggedleft 0.00 \tiny{(0.00)}} & \multirow{4}{\tempwidth}{\raggedleft 0.00 \tiny{(0.00)}} & \multirow{4}{\tempwidth}{\raggedleft 0.00 \tiny{(0.00)}} & \multirow{4}{\tempwidth}{\raggedleft 0.00 \tiny{(0.00)}}
			& \multirow{4}{\tempwidth}{\raggedleft 0.00 \tiny{(0.00)}} & \multirow{4}{\tempwidth}{\raggedleft 0.00 \tiny{(0.00)}} & \multirow{4}{\tempwidth}{\raggedleft 0.00 \tiny{(0.00)}} & \multirow{4}{\tempwidth}{\raggedleft 0.00 \tiny{(0.00)}} \\
			\hline
			\raggedright Depth of first detected failure
			& \multirow{2}{\tempwidth}{\raggedleft 8.56 \tiny{(22.82)}} & \multirow{2}{\tempwidth}{\raggedleft 17.50 \tiny{(62.33)}} & \multirow{2}{\tempwidth}{\raggedleft 7.55 \tiny{(9.12)}} & \multirow{2}{\tempwidth}{\raggedleft 44.04 \tiny{(109.78)}} & \multirow{2}{\tempwidth}{\raggedleft 30.87 \tiny{(74.51)}} & \multirow{2}{\tempwidth}{\raggedleft 24.28 \tiny{(75.26)}}
			& \multirow{2}{\tempwidth}{\raggedleft 3.85 \tiny{(0.49)}} & \multirow{2}{\tempwidth}{\raggedleft 4.66 \tiny{(1.64)}} & \multirow{2}{\tempwidth}{\raggedleft 9.96 \tiny{(14.38)}} & \multirow{2}{\tempwidth}{\raggedleft 13.20 \tiny{(16.06)}} \\
			\hline
			\raggedright Depth of first detected failure (gray box)
			& \multirow{3}{\tempwidth}{\raggedleft 8.56 \tiny{(22.82)}} & \multirow{3}{\tempwidth}{\raggedleft 17.63 \tiny{(62.61)}} & \multirow{3}{\tempwidth}{\raggedleft 7.55 \tiny{(9.12)}} & \multirow{3}{\tempwidth}{\raggedleft 44.04 \tiny{(109.78)}} & \multirow{3}{\tempwidth}{\raggedleft 30.87 \tiny{(74.51)}} & \multirow{3}{\tempwidth}{\raggedleft 24.28 \tiny{(75.26)}}
			& \multirow{3}{\tempwidth}{\raggedleft 3.85 \tiny{(0.49)}} & \multirow{3}{\tempwidth}{\raggedleft 4.66 \tiny{(1.64)}} & \multirow{3}{\tempwidth}{\raggedleft 9.96 \tiny{(14.38)}} & \multirow{3}{\tempwidth}{\raggedleft 13.20 \tiny{(16.06)}} \\
			\hline
			\raggedright Depth of first detected failure (black box)
			& \multirow{3}{\tempwidth}{\raggedleft 8.77 \tiny{(23.30)}} & \multirow{3}{\tempwidth}{\raggedleft 13.64 \tiny{(46.08)}} & \multirow{3}{\tempwidth}{\raggedleft 6.08 \tiny{(5.12)}} & \multirow{3}{\tempwidth}{\raggedleft 44.04 \tiny{(109.78)}} & \multirow{3}{\tempwidth}{\raggedleft 90.27 \tiny{(173.32)}} & \multirow{3}{\tempwidth}{\raggedleft 22.83 \tiny{(71.90)}}
			& \multirow{3}{*}{N/A} & \multirow{3}{*}{N/A} & \multirow{3}{*}{N/A} & \multirow{3}{*}{N/A} \\
		\end{tabularx}
		}
	\caption{Statistical data concerning the number of agents, the number of agent groups, the number of test cases, the number of reorganizations, as well as the occurrence, detection, and depth of failures. All undetected failures (see ``\%Undetected failures per test sequence'') can be attributed to error masking. ``\#Failures'' and ``\#Failures per test sequence'' refer to the number of faulty intermediate states the corresponding SO algorithm entered (note that this information is provided by our fault injection mechanism and not by the oracle that can only check for the validity of final states, i.e., reorganization results). All values are averages over the 700 generated test sequences per injected fault type. Values in parentheses denote standard deviations.}
	\label{tab:evaluation-results-faults}
	\vspace{-0.8cm}
\end{table}

The relatively high number of test sequences in which no failure was detected (between 70.22\% and 97.86\% on average) highlights the need for directed testing that is able to deal with the huge search space in a more efficient way (C-BranchingStateSpace).
For \psosp, the different numbers of applied test cases per test sequence reflect the difficulty of disclosing a specific type of fault using the black-box view.
In case of SPADA, all test cases were applied since the injected faults cannot be detected using the black-box view.
Another indicator for the difficulty of finding a specific fault type is the depth of the first detected failure (i.e., the index of the test case in which the first failure was detected).
Here, we see significant differences among the different SPADA and \psosp\ fault types.
This is because the system not only has to be pushed into a faulty intermediate state but the reorganization also has to end in a faulty state that can be detected by the oracle.
This challenge becomes clearer when taking a look at the percentage of undetected failures due to error masking, which ranges between 17.69\% and 97.75\% for {\psosp} and between 0.00\% and 99.84\% for SPADA.
Together with the average number of reorganizations per test sequence---that, except for SPADA-F3, significantly outmatches the average number of failures per test sequence---these observations illustrate the difficulty of testing~SOAS.

We observed that especially \psosp-F5 had a relatively high number of executed test sequences without detected failures (97.86\% compared to an average of 88.67\% over all other types of fault).
We therefore decided to use \psosp-F5 to investigate the influence of directed testing on \framework's ability to disclose a fault.
To study this effect, we introduced an additional fault type \psosp-F5d that is equivalent to \psosp-F5 but uses a specific system configuration:
To increase the chance of applying \psosp's approach exchange operator (an operation that is usually only applied in rare cases), we set the allowed number of partitions in \psosp-F5d to $\minNbPartitions = \maxNbPartitions$.
Using this parametrization, {\psosp} has no choice but to apply the random exchange or approach exchange operator, because
the split/join operators increase/decrease the number of partitions by one.
Note that this measure does not increase the code coverage (\psosp\ also applied the approach exchange operator in case of \psosp-F5), but the chance that the fault can be disclosed given the algorithm's non-deterministic behavior (C-BranchingStateSpace):
In approximately the same number of executed test cases, \psosp-F5d entered a faulty intermediate state about 23~times more often than \psosp-F5.
The benefit of directed testing is further reflected in an increase of the percentage of a faulty end state in case of a faulty intermediate state from 50.00\% for \psosp-F5 to 59.04\% for \psosp-F5d.
Consequently, the number of applied test sequences without detected failures dropped from $97.86\%$~to~$70.22\%$.

Summarizing, although \framework\ was able to find every type of injected fault, the high number of needed test sequences for failure detection demonstrates that the automatic generation of test suites on the basis of EPs and influence functions is especially useful for testing SOAS.
This is mainly because of the non-deterministic behavior of SO algorithms (C-BranchingStateSpace).
Our evaluation results also show that a combination of model-based and random generation techniques is essential to efficiently and effectively detect specific types of fault.
In particular, the gray-box interface is an important feature to mitigate the effect of error masking (C-ErrorMask). 
However, we expect an additional white-box view to certainly further increase the potential of finding faults in SO algorithms.

\section{Related Work}
\label{sec:relatedwork}
The necessity of testing adaptive systems has been recognized both in the testing community~\cite{Nguyen2009,Padgham2013,Zhang2009MBTagent,Wotawa2012AdapitveSystemsTest} and in the community of adaptive systems~\cite{Fredericks2013TRT,puschel2013towards,Cheng2009SEAMSroadmap1,deLemos2013SEAMSroadmap2}. 
Run-time as well as design-time approaches have identified non-determinism and the emergent behavior as main challenges for testing adaptive systems.

Run-time approaches for testing take up the paradigm of run-time verification~\cite{Falcone2011RV,Leucker08abrief,Filieri2012contREassurance}.
They shift testing into run-time to be able to observe and test, e.g., the adaptation to new situations.
Camara et al.~\cite{Camara2012ResilienceAdaptiveSystems} are using these concepts to consider fully integrated systems.
Their testing approach focuses mainly on testing non-functional properties of the system or more precisely, the resilience of the adaptive system.
The authors therefore investigate the system's adaptive capabilities by collecting and analyzing data in a simulated environment. The gained information is used as feedback for the running system. 
A similar approach is taken by Ramirez et al.~\cite{Ramirez2011ExploringUncertainty}, also focusing on non-functional requirements. 
The authors use the sampled data from a simulation to calculate a distance to expected values derived from the goal specification of the system. 
This information is used to adapt the system or its requirements proactively during run-time. 
Run-time approaches, however, are limited to tests of the fully integrated system and therefore are faced with problems like error masking which is very likely in such self-healing systems. 
In our layered testing approach---where {\framework} is placed on the interaction layer---we benefit from the piecemeal integration of the system for testing.
Thus, it is possible to avoid error masking within {\framework} by testing the SO algorithms isolated.
%Furthermore, the test systems have timing and performance constraints, because they are running in parallel to the actual system which should not be affected negatively by the testing activity. 

%but in our overall approach~\cite{Eberhardinger2014TeSOS} we follow a bottom-up testing strategy, starting from single agents, in order to cover testing in the whole software development life cycle. 
An important difference to the mentioned work is that we use these techniques for finding failures instead of analyzing the current system state for generating feedback for the adaptation. 
Still, we also use the basic concepts of run-time testing. 
The CEI allows us to split the evaluation into the three responsibilities R-Detect, R-Solution, and R-Distribution which in turn enable us to evaluate the runs without the evaluation of complex system states on the system level. 
As our evaluation shows, the CEI-based testing approach is especially beneficial in the context of self-organization. 

%Our presented framework is an important part of the entire testing approach~\cite{Eberhardinger2014TeSOS}, which is focused on self-organization in addition to adaptation. 
%
%Design-time approaches for testing adaptive systems~\cite{Nguyen2009,Padgham2013,Zhang2009MBTagent} focus on specific subsets, e.g., testing the implementation of agents separately. Zhang et al.~\cite{Zhang2009MBTagent} are testing the execution of plans of single agents and groups of agents. 
%This focal point does not allow to evaluate adaptive or even self-organizing characteristics of
%the system, contrary to our approach.

%In conclusion, testing adaptive systems is in the focus of some research groups.  
%To our knowledge there is no approach which is extending the techniques to self-organizing (SO) systems. We are able to deal with SO systems in this paper and show how it is possible to test algorithms which are responsible for reconfiguring the structure and organization of systems during run-time.
%
%Our presented approach on testing self-organization algorithms is mainly formed by the ideas on probabilistic modeling the environment with environment profiles (EPs) as well as the presented concepts of the isolated framework;
%Next we shortly summarize the related work in this two areas.

Design-time approaches like~\cite{Stott2000FaultInjection,Nguyen2009,Padgham2013,Zhang2009MBTagent,luckey_eval4saso12} test the systems in a classical manner during the development.
All these approaches are considering some dedicated parts of the system.
Consequently, it is not possible to give evidence about the correct functionality of the overall system.
Zhang et al.~\cite{Zhang2009MBTagent} compose their tests towards a fully integrated system test, but they do not consider adaptivity or SO explicitly since they focus on testing the correct execution of plans within multi-agent systems. 
Nguyen et al.~\cite{Nguyen2009} promote an approach for a component test suite (where the components correspond to agents), but do not consider interaction or organization between the agents as it would be necessary for SO.

The evaluation of the test results, i.e., the application of a test oracle for adaptive behavior is only considered by Fredericks et al.~\cite{Fredericks2013TRT,Fredericks2014RunTimeTestingUncertainty} and Nguyen et al.~\cite{Nguyen2013}. 
Both approaches are relying on goals reflecting the requirements of the system that are somewhat loosened in order to reflect the ever-changing environment the agents have to adapt to: 
The approaches mitigate the goals with the \emph{RELAXed} approach~\cite{DBLP:conf/re/WhittleSBCB09} or consider soft goals that do not need to hold at all time. 
Consequently, the decision of the test oracle is rather fuzzy. 
In our approach the definition of correct and in-correct behavior is given by the CCB that enables us to decide whether a failure occurs or not. 

\paragraph{Probabilistic Test Models of Environmental Changes and Influences.}
There are several approaches to tackle uncertainties about the expected usage of a SuT within testing models. They can be summed up under the idea of \emph{operational profiles}~\cite{Smidts2014TestingWithOperationalProfile}. 
The information within these test models represents the user's behavior in a probabilistic model. 
For this purpose, different techniques for generating and using these profiles have been provided.
 
Sammodi et al.~\cite{Sammodi2011UsageBasedOnlineTesting}, e.g., generate usage profiles, as they call them, by monitoring the user's interaction with the system and deriving the profiles for the observed usage afterward. 
One of our possibilities to establish EPs also follows this monitoring and analysis process with the difference that we are not monitoring users of the system. Instead we monitor the whole system environment which includes all influences on the SOuT. 
Another approach to design models of the usage is presented by Samih et al.~\cite{Samih2014UsageModels}. 
They enrich the models by introducing capabilities in order to model variants of specific features, i.e., product features, to form test models for product line engineering. 
The approach made by Ehlers et al.~\cite{Ehlers2011SASMonitoringWithUsageProfile} focuses on using the usage profiles for detecting anomalies in adaptive systems in order to use the information about the anomalies during the adaptation process.  
Besides focusing on handling user behavior, there is also some work on representing the behavior of other system components in the test model, like Popovic et al.~\cite{Popovic2007RobustnessTesting}. 
The authors use the models for protocol testing and therefore represent valid and invalid communication between components.

Operational profiles thus are an established technique for modeling uncertain behavior---mainly of the user---for designing test models and for evaluation purposes. 
In {\framework}, we use environment profiles which are based on a similar concept to deal with the complexity of the ever-changing environment of SOAS by reducing its state space using a probabilistic approach. 
%Our profiles are used for abstracting from concrete environment states to achieve isolated testing for one SO algorithm (C-Isolate). 
% Using our functions $f_\mathcal{G}$ describing the environment's influence enables us to control the size of the state space (C-BranchingStateSpace).

\paragraph{Isolated Testing of Component-Based Systems.}
Our presented concepts for a framework for isolated testing of SO algorithms are related to methods and techniques in the research area of isolated testing of component-based systems. 
A recent approach in this area by Thillen et al.~\cite{Winkler2014IsolatedTestingDistributedSystems} promotes the ``tester in the middle'' idea that aims at improving testing of distributed components depending on other components in the system. 
Their application area is the testing of network components. 
For this purpose, they model dependencies within the network to be able to build mock-ups out of this models. 
Bauer et al.~\cite{Bauer2010EnablingStatisticalTestingComponent-basedSystems} propose a statistical strategy for isolated testing of component-based systems. Their approach is based on state-based models that are used to generate interaction test models. 
The goal is to test the interactions and functionalities within the system under test which is composed of several system components. 
%A more general approach for improving testing in component-based systems is made by Toroi~\cite{Toroi2009ConformanceTestingComponentBasesSystems}. 
%The aim of his work is to enhance testing methods from the integrator view, i.e., to handle the composition of the. 
Yao and Wang~\cite{yao2005framework} present a framework for testing distributed software components that provides an environment to allow a client-side software component to define tests for a black-box component published on the server-side.
The framework focuses on automatic test execution without considering explicitly the generation and evaluation of the tests. 
Wu et al.~\cite{Wu2008Jata} propose \textsc{Jata}, a testing language for distributed components enabling the use \textsc{JUnit} in the context of service-oriented systems and offering support for a message-oriented middleware; like Yao and Wang~\cite{yao2005framework} it focuses on the execution of component tests on web services.
In contrast to the approaches in~\cite{yao2005framework,Wu2008Jata}, {\framework} offers a complete approach from test generation over execution to test evaluation with a focus on agent-based systems and functional testing of single SO algorithms.

\smallskip%
Our approach for isolated testing of SO algorithms is an efficient combination of model-based techniques using the concepts of isolated testing and probabilistic modeling in order to tackle the challenges of testing SOAS.
To our knowledge, there is no approach extending both of these techniques to SO algorithms.
%In SAOS, we cannot assume to have a stable system structure and have to cope with agents entering and leaving the system within a test run.
%However, the idea of using state-based models as input for the test suite generator -- in order to abstract from the other components -- is very similar.

\section{Conclusion and Future Work}
\label{sec:conclusion}

We motivated the necessity of testing SOAS and outlined an approach that copes with the complexity arising from the characteristics of these systems.
We introduced the framework {\framework} for testing SO algorithms that is an important element coping up with this challenge.
To be able to test SO algorithms in a systematic and automatic fashion, we identified four challenges that are addressed: error masking among the algorithms of SOAS, interleaved feedback loops which distort test results, the oracle problem in the context of SO, and the ramified state space spanned by SO algorithms. 
We were able to show that it is possible to meet these challenges by isolating the SO algorithms and applying a model-based testing approach. 
The resulting framework encompasses automatic test suite generation, execution, and evaluation within a controlled environment. 
Environment profiles are used to abstract from concrete environments and allow for generating large test sequences. 
This technique showed to be extremely valuable since a lot of test runs are necessary to find a trace through a SOAS that actually reveals a failure.
 
We evaluated {\framework} on the basis of two different partitioning-based SO algorithms in an existing smart-grid application. 
Our results demonstrate that an efficient combination of model-based and random test case generation techniques allows to find different kinds of failures in an acceptable time limit. 
In particular, the gray-box view turned out to be very important to reduce the effect of error masking. 

\paragraph{Future work} focuses on three aspects: first, \emph{extending the evaluation} of the test run results enabling \emph{fault diagnosis}; second, gaining more \emph{control on the execution} to increase \emph{reproducibility} of the test results; and third, using {\framework} to investigate non-functional aspects of the system, i.e., to evaluate the performance of SO-algorithms.  

For the extension toward fault diagnosis, we aim at combining the different test sequences of a specific test suite into a \emph{test sequence evaluation tree}. 
For this purpose, we will try to identify equal prefixes of the test sequences. 
If two or more test sequences share a common prefix, they will share a common path in the tree.
Thus, it is possible to unfold overlaps between executed and evaluated test sequences and map them into a tree model. 
This tree could be used for fault localization, optimization of later test generations, or visualization of the results for a deeper understanding. 

%As already mentioned in our evaluation (cf.\ \cref{sec:evaluation}), SOAS suffer from a vast number of possible traces and the possibility of error masking. To further minimize or even exclude error masking (C-ErrorMask), we need, additionally to the black- and gray-box testing, a \emph{white-box view}. This requires the possibility of sampling interim results of every step the SO Algorithm takes during test case execution.
%In such a setting, the problem will be the huge amount of sampled data to analyze and the loss of generality of the approach.
%The latter is because a white-box view necessitates algorithm-specific knowledge.

%Furthermore, we are going to extend the framework in order to measure the quality of a SO algorithm. 
%Therefore, we follow the hypothesis that a system within the corridor of correct behavior performs better, i.e., has a higher quality, than one outside, which is true by the definition of the corridor.
%Following, it would be of interest to measure the overall time of the system controlled by a specific SO algorithm being within and vice versa the time outside the corridor.
%The results should be used as a first step to extending the framework towards measuring quality.

To increase the \emph{reproducibility} of the test results, the test system has to ensure the following properties: (1)~a random seed has to be used in the whole test system; (2)~the scheduling of the underlying execution platform has to be controlled in a reproducible way; and (3)~the concurrent execution of a distributed test system has to be synchronized in a deterministic way.
There are different techniques to address (1) to (3) in order to detect so-called ``Mandelbugs''\footnote{According to Grottke and Trivedi~\cite{Grottke2005ClassificationOfFaults} a Mandelbug is ``[a] fault whose activation and/or error propagation are complex, where `complexity' can take [the following form]: [\ldots] The activation and/or error propagation depend on interactions between conditions occurring inside the application and conditions that accrue within the system-internal environment of the application [\ldots]''.} (cf.\ the \textsc{Chess} tool by Musuvathi et al.~\cite{Musuvathi2008FRH} or the empirical study on this topic by Thomson et al.~\cite{Thomson2014CTU}). These could be used to provide an adequate underlying platform for {\framework}.
Although designing such a platform was not in the scope of this paper, we met the points~(1) and partially~(3).
As we perform our tests within the limits of the underlying execution platform (Java), which does not affect the ability to detect faults but might affect the reproducibility (at least in some cases), we currently do not meet~(2) and achieve~(3) only on the basis of a stepwise execution model.
We are going to extend our experiments with a continuous execution model to evaluate differences in the test results and its effects on the reproducibility as well as the ability of {\framework} to execute tests in this environment. 
% Further, it is however still an open question to which degree the synchronization of concurrent executions of distributed test is possible in a deterministic way.

Besides investigating functional tests we already started to use {\framework} for the evaluation of performance criteria. 
In~\cite{eberhardinger2015research}, we showed its ability to perform evaluations on non-functional aspects of SO algorithms and developed a set of five major requirements for metrics that evaluate the performance of SO algorithm within {\framework}.
The development of the metrics and the integration into the framework is a further topic of future work in testing self-organizing, adaptive systems.

\paragraph{Acknowledgment.}
This research is sponsored by the research project \emph{Testing Self"=Organizing, adaptive Systems (TeSOS)} of the German Research Foundation.

\bibliographystyle{splncs03}
\bibliography{testingSOalgorithms}

\end{document}